 \documentclass[final,5p,times,twocolumn]{elsarticle}
\usepackage[colorlinks=true, allcolors=blue]{hyperref}
\usepackage{amssymb}
\usepackage{lipsum}
\usepackage{subcaption}
\usepackage{amsmath}
\usepackage{bm}

\journal{Physics Letters B}

\begin{document}

\begin{frontmatter}



\title{In situ enhanced detectability of axions' responses in a microwave cavity}


\author[first]{Li Gao}
\author[first]{Suirong He} 
 \author[first]{Hao Zheng}
  \author[first]{Ni Chen }
   \author[first]{Xianing Feng}
   \author[second]{Qingquan Jiang}
   \author[third]{ Jin Li}
 \author[first]{L. F. Wei}%

\affiliation[first]{organization={Southwest Jiaotong University},
            addressline={HgerD collaboration and Information Quantum Technology Laboratory, School of Information Science and Technology}, 
            city={Chengdu},
            postcode={610031}, 
            state={Sichuan},
            country={China}}

\affiliation[second]{organization={China West Normal University},
            addressline={HgerD Collaboration and School of Physics and Astronomy}, 
            city={Nanchong},
            postcode={637002}, 
            state={Sichuan},
            country={China}}

\affiliation[third]{organization={Chongqing University},
            addressline={HgerD Collaboration and School of Physics}, 
            city={Chongqing},
            postcode={401331}, 
            state={Chongqing},
            country={China}}

\begin{abstract}
Microwave cavities in strong magnetic fields are among the most sensitive detectors for axion dark matter detection. Its sensitivity, however, is fundamentally limited by the second-order nature of axion-photon energy conversion, which yields an extremely weak signal. Here we propose a feasible approach to break this limitation, by pre‑exciting the target cavity mode to generate a first-order power response signal in situ. The resulting signal amplitude scales linearly with the axion–photon coupling parameter, which is in sharp contrast to the quadratic scaling of conventional schemes, and is detectable via standard quadrature balanced homodyne detection system, even though its phase remains stochastic. Numerical simulations demonstrate that this pre-excited seed field can effectively suppress thermal noise, and thus the detection sensitivity of the system can approach the lower quantum noise limit. Consequently, the sensitivities predicted by the KSVZ and DFSZ models could be surpassed. This proposal might offer a pathway to a major leap in axion dark matter searches, as it is within reach of current technology.
\end{abstract}




\end{frontmatter}




\section{Introduction}
\label{introduction}
The existence of dark matter is one of the most profound unsolved problems in modern physics~\cite{DarkF}. An axion is a pseudo-scalar particle originally proposed to resolve the strong CP problem in QCD and now is regarded as the most promising dark matter candidates.
Based on cosmological abundance observations, the axion decay constant $f_a$ is predicted to lie in the range $10^9$ \text{GeV} $\sim 10^{12}$ \text{GeV}, corresponding to the mass $m_a$
  in the \text{$\mu$eV} $\sim$ \text{meV} regime~\cite{KSVZ1,KSVZ2,DFSZ1,DFSZ2}. Physically, the detection of axions is not only directly relevant to verifying Peccei–Quinn (PQ) symmetry, but also intimately connected with gravitational wave physics, thus attracting intense interest~\cite{HXZ}.

Various detection schemes have been developed for axion searches, typically including the lumped-element experiment DMRadio-$\text{m}^3$ covering the 5 \text{MHz} $\sim$ 200 \text{MHz} band~\cite{DMR}, CASPEr working at 30 \text{MHz} ~\cite{CASPEr,CASPEr1}, the nuclear magnetic resonance (NMR) measurement experiment in the 400 \text{MHz} $\sim$ 500 \text{MHz} range~\cite{KRbNe}, QUAX in the 10 \text{GHz} band~\cite{QUAX,QUAX1}, and CADEx at higher frequencies of 80 \text{GHz} $\sim$ 110 \text{GHz}~\cite{CADEx}, etc.
Particularly, Haloscope-type experiments, which exploit the inverse Primakoff effects in strong magnetic fields~\cite{Primakoff1,Primakoff2,Primakoff} for the electromagnetic response detection of axions in the 500 \text{MHz} $\sim$ 10 \text{GHz} band~\cite{QCP}, have attracted extensive research interest. The representative experiments include, e.g., ADMX~\cite{ADMX}, CAPP~\cite{CAPP,CAPP1}, LSW-ALPSI~\cite{ALPSI}, ALP~\cite{ALP}, CARRACK~\cite{CARRACK}, CULTASK~\cite{CULTASK}, ABRACADABRA~\cite{ABRACADABRA}, TOORAD~\cite{TOORAD}, and also ORCT~\cite{ORCT}, etc. Remarkably, their achievable detection sensitivities have been approaching the levels predicted by the KSVZ and DFSZ axion models. Nonetheless, to date, no evidence for the existence of axions has been found in these experiments, although they are operated under ultralow temperature for the extremely high‑sensitive detection. Therefore, how to further improve the achievable sensitivity by enhancing its detectability has become an urgent challenge.

Notably, the signals detected in almost all existing Primakoff-effect-based Haloscope-type experiments are the second-order energy responses, since the target cavity mode is maintained at its ground state. As a consequence, the detectable power of the response signal scales with $g_{a\gamma\gamma}^2$, making such signals inherently difficult to detect.
To overcome this challenge, and building on previous work~\cite{GL}, we generalize the one-dimensional model to a three-dimensional cylindrical cavity model and explicitly specify the target mode pre-excited in the microwave cavity, namely, the \(\mathrm{TM}_{010}\) mode. This pre-excited seed field significantly amplifies the axion-induced response signal in situ. As a result, a first-order power signal is produced whose amplitude scales linearly with the axion–photon coupling constant $g_{a\gamma\gamma}$, in contrast to conventional second-order responses that depend on $g_{a\gamma\gamma}^2$.  Furthermore, we present a rigorous analysis of the signal and noise generation and propagation mechanisms along the entire detection and transmission chain. Relative to previous work, this analysis additionally incorporates the phase noise and amplitude noise of the pump signal, as well as the noise in the transmission chain. Numerical noise estimates indicate that the pump-induced phase noise, amplitude noise, and shot noise substantially exceed the transmission-chain noise and thermal noise. Accordingly, we examine the influence of the pump-signal strength on the sensitivity enhancement and thermal-noise suppression, and thereby establish a constraint on the pump-signal strength.

 Importantly, the amplitude of such a first-order power signal is detectable for the usual quadrature balanced homodyne detection technique, although its phase might still be stochastic. Benefiting from such a significant enhancement of detectability, we show that the achievable detection sensitivity can be improved by roughly 1 orders of magnitude over the existing Haloscope-type experiments wherein only the second-order energy response signal can be detected. Strikingly, the pre-excited seed field strongly suppresses the thermal noise in such an upgraded detector, permitting ultrahigh target detection sensitivity at room temperature.

\section{Model and Method.}\label{Sec.2}
\begin{figure}
    \centering
    \includegraphics[width=0.8\linewidth, height=5cm]{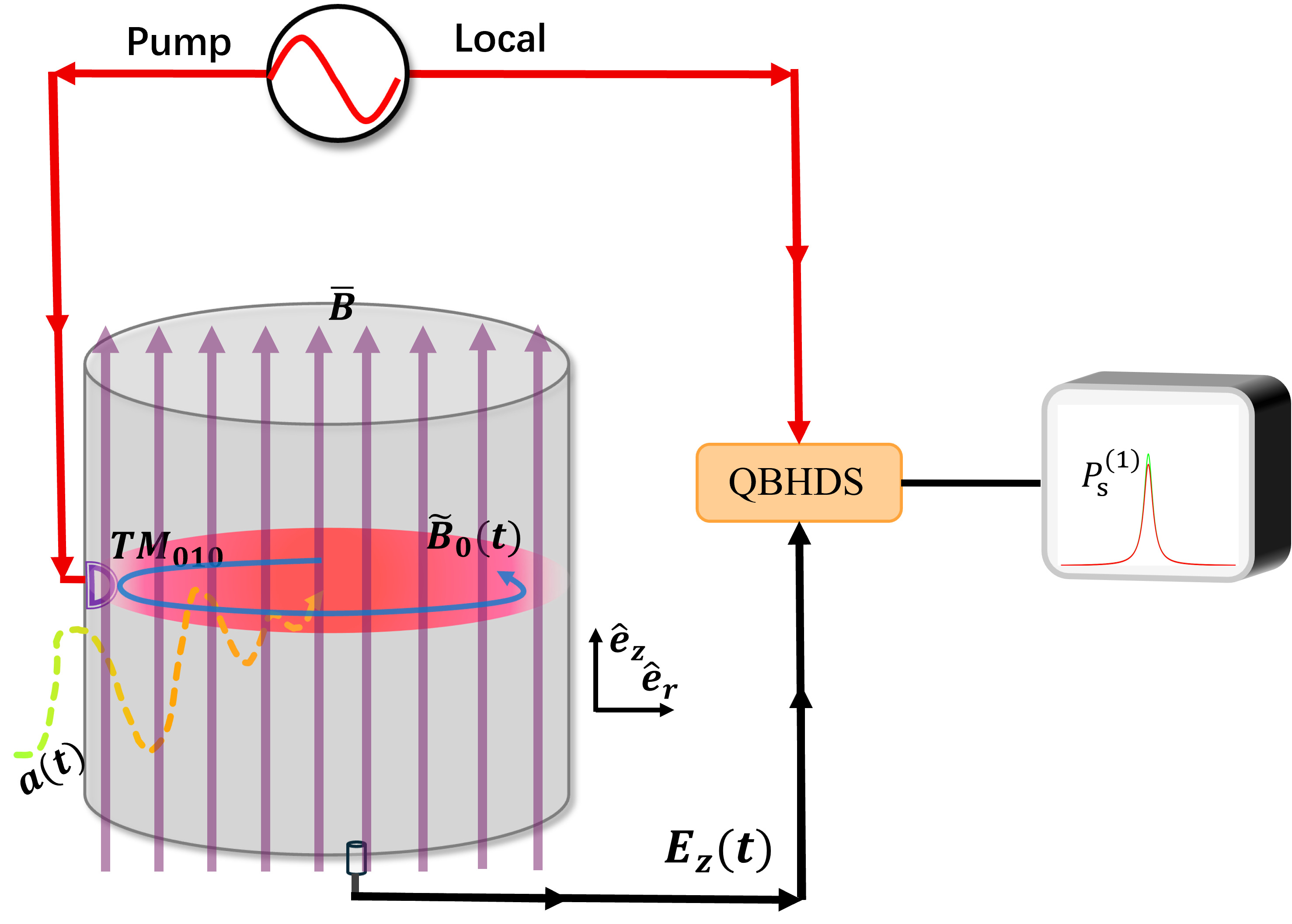}
    \caption{Schematic of the upgraded axion-detection setup incorporating a microwave resonant cavity. Local oscillator signals and intracavity fields $\bm{E}_z(t)$ feed into the QBHDS (Quadrature Balanced Homodyne Detection System) circuitry, enabling efficient extraction of the amplitude $P_s^{(1)}$ of the generated first-order power signal.}
    \label{fig1}
\end{figure}

The upgraded configuration of the conventional Haloscope-type detector is schematically illustrated in Fig.~\ref{fig1}, where a strong static magnetic field $\bar{B}$
  is established along $\bm{e}_z$ inside the cylindrical cavity. Notably, within this configuration, a magnetically coupled loop antenna is placed at the mid-height of the cavity, thereby achieving a steady-state pre-excitation of the $\mathrm{TM}_{010}$
  mode~\cite{YZQJ}.
The axion-induced response signal is subsequently superposed onto this seed field to generate a first-order power signal, where the axion signal is enhanced in situ. The amplitude of the generated first-order power signal (although its phase might still be stochastic) can be extracted via the usual quadrature balanced homodyne detection system. As the amplitude of this first-order power signal varies linearly with $g_{a\gamma\gamma}$, both the detectability of axion electromagnetic response signals and the ultimate achievable detection sensitivity are markedly improved.

In the International System of Units, the Lagrangian density that governs the axion‑photon interaction inside a microwave cavity can be expressed as~\cite{Skivie}:
\begin{align}\label{1}
    \mathcal{L}=&-\frac{1}{4\mu_0}\mathcal{F}_{\mu\nu}\mathcal{F}^{\mu\nu}+\frac{1}{2\hbar c}[\partial_\mu a\partial^\mu a-\frac{m_a^2c^4}{\hbar^2c^2}a^2]\nonumber\\
   & -\frac{g_{a\gamma\gamma}}{4\mu_0}a\mathcal{F}_{\mu\nu}\tilde{\mathcal{F}}^{\mu\nu}-J^\mu A_\mu,
\end{align}
where 
$\mathcal{F}_{\mu\nu}$ and $\tilde{\mathcal{F}}^{\mu\nu}$ denote the electromagnetic field tensor and its covariant form, while $a$ and $m_a$ represent the axion field and its mass, respectively. 
$g_{a\gamma\gamma}=g_\gamma\alpha/(\pi f_a)$ is the axion–photon coupling coefficient and $g_\gamma=0.36 (-0.97)$ is predicted by the DFSZ~\cite{DFSZ1,DFSZ2} ( KSVZ~\cite{KSVZ1,KSVZ2}) model.  Specifically, $f_a\simeq 10^{12}$ \text{GeV} stands for the axion decay constant, $\alpha=1/137$ is the fine-structure constant, $\hbar$ denotes the reduced Planck constant, and $\mu_0$ the vacuum permeability. $J^\mu$ and $A_\mu$ are the external source and the electromagnetic potential, respectively. 
With Eq.~\eqref{1}, the axion-modified Maxwell equations can be obtained as:
\begin{equation}\label{2}
  \left\{
    \begin{aligned}
        &\frac{1}{c^2}\partial_t\bm{E}+\nabla\times\bm{B}=\mu_0(\bm{J}+\bm{J}_{eff}),\\
    &\nabla\cdot \bm{E}=\frac{1}{\varepsilon_0}(\rho+\rho_{eff}),
    \end{aligned}
    \right.
\end{equation}
where $\varepsilon_0$ is the vacuum permittivity, $\bm{E}$ and $\bm{B}$ stand for the electric and magnetic fields inside the cavity, respectively. Moreover, $\rho$ represents the free charge and $\bm{J}(\bm{x})$ denotes the current density inside the cavity. Given that the axion field can be treated as approximately isotropic and homogeneous, i.e., $\nabla a\sim 0$, the effective charge induced by axion effects is negligible, i.e., $\rho_{eff}\sim 0$. Ultimately, the effective current $\bm{J}_{eff}=-g_{a\gamma\gamma}\bm{B}^{(0)}\partial_ta/({4\mu_0c})$ describes the electromagnetic response of axion.
Importantly, in conventional Haloscope-type experiments~\cite{ADMX,CAPP,CAPP1,ALPSI,ALP}, the cavity is generally operated in a passive source‑-free state with $\bm{J}=0$. This implies that the detectable signal is just generated by the axion‑induced excitation of the electromagnetic vacuum within the cavity. The detectable power is thereby proportional to the square of the axion–photon coupling coefficient $g_{a\gamma\gamma}$, and thus it is particularly weak. By contrast, in the present configuration $\bm{J}\neq 0$, the electromagnetic field strength inside the cavity can be expanded as
\begin{equation}\label{3}
    \left\{
\begin{aligned}
    \bm{E}=\bm{E}^{(0)}(g_{a\gamma\gamma}^0)+\bm{E}^{(1)}(g_{a\gamma\gamma})+\mathcal{O}(g_{a\gamma\gamma}^2),\\[3pt]
    \bm{B}=\bm{B}^{(0)}(g_{a\gamma\gamma}^0)+\bm{B}^{(1)}(g_{a\gamma\gamma})+\mathcal{O}(g_{a\gamma\gamma}^2),
\end{aligned}
\right.
\end{equation}
where $\bm{E}^{(0)}$ and $\bm{B}^{(0)}$ denote the pre‑existing seed field without any axion response (called axion’s zeroth-order electromagnetic response). 
For the pre-excited $\mathrm{TM}_{010}$ mode $\omega_m=\omega_{010}$, the axion’s zeroth‑order response obeys the following driven dissipative equation:
\begin{align}\label{4}
    &(\frac{d^2}{dt^2}+\gamma_m\frac{d}{dt}+\omega_m^2)\tilde{B}_0(t)=c^2\mu_0C_{0}J_0\bm{X}_m(\bm{x})\cos{(\omega_0t)},
\end{align}
where $C_{0}=\int_V\bm{X}_m^*(\bm{x})\cdot\nabla\times\bm{J}(\bm{x})dV$ is the overlap coefficient between $\bm{J}(\bm{x},t)$ and normalized spatial profile $\bm{X}_m(\bm{x})$ of the $\mathrm{TM}_{010}$ mode.  Furthermore, $\gamma_m=\omega_{m}/Q$ denotes the dissipation coefficient of the $\mathrm{TM}_{010}$
  cavity mode with $Q$ denoting its quality factor. The steady-state solution to Eq.~\eqref{4} reads
\begin{align}\label{5}
\tilde{B}_0(\bm{x},t)=\frac{c^2\mu_0J_0C_0k_c\bm{X}_m(\bm{x})}{\omega_m^2-\omega_0^2-i\omega_0\gamma_m}\cos{(\omega_0t)},
\end{align}
where $k_c=x_0^1/R$  
with $x_0^1$ being the first root of the zero-order Bessel function and $R$
the cavity radius. The total magnetic field strength of the axion’s zeroth‑order electromagnetic response inside the cavity is given by $ \bm{B}^{(0)}(\bm{x},t)=\tilde{B}_0(\bm{x},t)\bm{e}_\theta+\bar{B}(\bm{x})\bm{e}_z$.
Using the fundamental electromagnetic relation $\nabla\times\bm{\tilde{E}}_{0}=-\partial_t\bm{\tilde{B}}_{0}$, the corresponding steady-state electric field $\tilde{E}_0(\bm{x},t)$
 can be derived. 

The core innovation lies in preparing the target cavity mode in a steady-state excited field $\tilde{B}_0$, rather than in the vacuum state, and the field magnitude can be flexibly tuned by adjusting the amplitude of the excitation source. Specifically, the second term in Eq.~\eqref{3} describes the first-order electromagnetic response induced by axions via the Primakoff effect, and this response obeys the axion effective current-driven electromagnetic field equation below:
\begin{equation}\label{6}
\nabla\times\bm{B}^{(1)}+\frac{1}{c^2}\partial_t\boldsymbol{E}^{(1)}=-\frac{1}{4c}g_{a\gamma\gamma}\bm{B}^{(0)}\partial_ta,
\end{equation}
where the general form of the axion field reads $a(\bm{x},t)=\mathrm{Re}[a_0{ \exp}({i\bm{k}_a\cdot\bm{x}-i\omega_a t-i\phi_a)}]\sim a_0\cos{(\omega_a t+\phi_a)}$ with $a_0=(2\rho_a\hbar c^5/\omega_a^2)^{1/2}$, $\omega_a$ and $\bm{k}_a$ being the amplitude, frequency, and wave vector of the axion field, respectively.  $\rho_a=0.5\times10^{-24}$ $\mathrm{g\cdot cm^{-3}}$ is the dark matter axion density and $\phi_a$ is the axion field phase that varies randomly in $[0,2\pi]$ for times longer than the axion coherence time $\tau_a\sim 2\pi \hbar c^2/(m_a c^2\sigma_v^2)$ with velocity dispersion $~\sigma_v=10^{-3}c$~\cite{taua}. 
Within the Lorenz gauge, Eq.~\eqref{6} can be rewritten as
\begin{align}\label{7}
    \left(\nabla^2-\frac{1}{c^2}\partial_t^2+ \gamma \partial_t\right)\bm{A}^{(1)}=\mu_0(J_{ez}\bm{e}_z+J_{e\theta}\bm{e}_\theta),
\end{align}
where $J_{ez}=-g_{a\gamma\gamma}\bar{B}\partial_ta/(4c\mu_0)$ represents the effective current identical to that in conventional Haloscope-type detector, $J_{e\theta}=-g_{a\gamma\gamma}\tilde{B}_0\partial_ta/(4c\mu_0)$ denotes another effective current arising from the coupling between the excited-mode magnetic field $\tilde{B}_0$ and the axion, and the dissipation coefficient $\gamma=\omega/Q$ is related to the excited cavity mode frequency $\omega$ and the cavity quality factor $Q$. 
Evidently, the steady-state response in the targeted $\mathrm{TM}_{010}$ mode resulting from $J_{ez}$ reads
\begin{align}\label{8}
   & E_{mz}^{(1)}(\bar{B},\omega_a, \phi_a, t)
    = a_0g_{a\gamma\gamma} C_1\bm{X}_{m}(\bm{x})  \frac{c\omega_a^{2}\bar{B}}{4D_m}\bm{e}_z\nonumber\\
        &\quad\times\left[\Gamma_m\sin{(\omega_at+\phi_a)}+\Delta_m\cos{(\omega_at+\phi_a)}\right],
\end{align}
where $D_m=\Delta_m^2+\Gamma_m^2$, with $\Delta_m=\omega_m^2-\omega_a^2$ and $\Gamma_m=\gamma_m\omega_a$, and $C_1=\int_V \bm{X}_m^*(\bm{x}) \cdot \bm{\bar{B}}(\bm{x}) dV$. 
It is interesting to note that the effective current $J_{e\theta}$ might excite the $\text{TE}_{011}$ mode. See the full derivation  provided in \ref{Appendix A}.
Consequently, the total electromagnetic field in the cavity can be expressed as:
$\bm{E}(\bm{x},t)=\bm{E}^{(0)}+\bm{E}^{(1)}=E_{l\theta}^{(1)}\bm{e}_\theta+(\tilde{E}_0+E_{mz}^{(1)})\bm{e}_z$ and $\bm{B}(\bm{x},t)=\bm{B}^{(0)}+\bm{B}^{(1)}=B_{l\rho}^{(1)}\bm{e}_\rho+(\tilde{B}_0+B_{m\theta}^{(1)})\bm{e}_\theta+(\bar{B}+B_{lz}^{(1)})\bm{e}_z$, respectively. The instantaneous signal power in the resonant cavity reads as: $ P_s(t)=\sum _{i=0}^{2}P_s^{(i)}(t)$, where
$
P_s^{(i)}(t)=P_s^{(i)}f_{i}(t)$ for $i=1,2,3$ denote the signal powers. Here $P_s^{(i)}$ are the amplitudes of the zeroth-, first- and second-order axion signals, and $f_i(t)$ is the instantaneous time-dependent part.
For the resonant responses with  $\tilde{B}_0\ll \bar{B}$,  $\omega_0=\omega_a=\omega_m$ and $\omega_+=\omega_l$, the detectable power contributed by the second-order electromagnetic response of the axion can be expressed as
\begin{align}\label{9}
    P_s^{(2)} \simeq\frac{1}{8\mu_0}a_0^2g_{a\gamma\gamma}^2\bar{B}^2\omega_aQC_1V,
\end{align}
This is just the signal power generated in the usual Haloscope-type experiments for detection (the detailed derivations are briefly given in \ref{Appendix B}). Accordingly, with the typical experimental parameters~\cite{P}, it can be estimated as $P_s^{(2)}\sim 3.63\times10^{-25}$ \text{W}, which could be detected only by using the  ultrahigh-sensitivity microwave single-photon detector, in principle. In comparison, the signal power originating from the first-order axion response (derived in \ref{Appendix B}) reads
\begin{align}\label{10}
&P_s^{(1)}(\phi_a,t)=P_s^{(1)}\cos{(2\omega_at+\phi_a)},
\end{align}
whose amplitude $P_s^{(1)}=a_0g_{a\gamma\gamma}\bar{B}\tilde{B}_0\omega_aVC_l/(8\mu_0)$ should be significantly large for detection (see \ref{Appendix B} and \ref{Appendix C}). Above, $C_l=C_1+C_2/4
$ is the coupling coefficient between the cavity mode and the signal.
Indeed, with the typical experimental parameters, $C_0=C_1=C_2=1$, $\omega_a=2\pi \rm ~GHz$, $V=\pi R^2H=5.13$ \text{L}, $\tilde{B}_0=3$ $n \text{T}$, $|\bar{B}|=8$ \text{T} and $Q=10^4$, it can be written as
\begin{align}\label{11}
&P_{s}^{(1)}=3.24\times10^{-17}\mathrm{W}\left(\frac{g_\gamma}{0.36}\right)\left(\frac{10^{12}\mathrm{GeV}}{f_a}\right)\left(\frac{\omega_a}{2\pi~\mathrm{GHz}}\right)\nonumber\\
&\times\left(\frac{2\rho_a}{ 10^{-21}{\rm kg\cdot m^{-3}}}\right)\left(\frac{\tilde{B}_0}{3\mathrm{n T}}\right)\left(\frac{\bar{B}}{8\mathrm{T}}\right)\left(\frac{Q}{10^4}\right)\left(\frac{V}{5.13\mathrm{L}}\right).
\end{align}
This indicates clearly that the amplitude of the first-order power, arising from the interference of the zero- and first-order responses,  numerically surpasses the second-order response power of the axion by about 8 orders of magnitude. Therefore, if this first-order power signal is detectable, the detectability of the axion response signal will be significantly enhanced, thereby substantially improving its achievable detection sensitivity.

\section{The achievable detection sensitivity.}
The first-order signal power in Eq.~\eqref{10} depends on the random phase $\phi_a$ of the axion field, leading to a null statistical average via traditional direct power detection, although it can still be amplified by using a low-noise quantum amplifier.
 However, \ref{Appendix C} demonstrates that the first-order signal amplitude can be retrieved by quadrature-balanced homodyne detection. After comprehensive noise analysis, we can derive the axion detection sensitivity based on the amplitude of the first-order electromagnetic response power.

 We known that $P_{s}^{(0)}\propto (\tilde{E}_{\rm 0,sys})^2$ is the zeroth-order power, $P_{s}^{(2)}\propto (E_{mz}^{(1)})^2$ is the axion second-order power, and $P_{s}^{(1)}=(P_{s}^{(0)}P_{s}^{(2)})^{1/2}\propto\tilde{E}_0E_{mz}^{(1)}$ is the first-order signal power. Since the second-order power is much smaller than the zeroth-order power, i.e., $P_{s}^{(2)}\ll P_{s}^{(0)}$, it can be neglected.
In \ref{appendix 4}, we provide a detailed analysis of the noise affecting the axion signal as it propagates from the cavity through the antenna, transmission chain, and double-balanced homodyne detection system to the detection terminal, together with the corresponding definition of the signal-to-noise ratio (SNR). The noise in this process mainly consists of pump-source phase noise $\sigma_{P,\phi}=|\frac{\partial P_s^{(0)}}{\partial\phi}|\sqrt{S_\phi\Delta f}$ (with the phase-noise PSD approximated as $S_\phi(f)=-100{\rm dB ~rad^2/Hz}$) and amplitude noise $\sigma_A=P_s^{(0)}\sqrt{{\rm RIN}\Delta f}$ ( with Relative Intensity Noise (RIN) as an external system parameter, and adopting  ${\rm RIN}=-120{\rm dB/Hz}$), shot noise $P_{\rm shot1}=\sqrt{P_s^{(0)}\hbar\omega_a/\tau}$ arising from photon-number fluctuations in the pump field, cavity thermal noise $P_{T,cav}=k_BT_{cav}\Delta f$, antenna-coupling noise $P_{N,ant}=k_BT_{N,ant}\Delta f$, and transmission-line noise $P_{N,line}=k_BT_{N,line}\Delta f$.
The axion-containing signal power received at the detection terminal is given by
\begin{align}
    P_{\rm out}^{s}=\frac{8\eta h\mathcal{R}^2P_L}{Z_D}(P_s^{(1)}+P_s^{(2)})
\end{align}
where $\eta=\beta/(1+\beta)$ denotes the antenna coupling efficiency, $Q_i=\beta Q_c$ is the external coupling quality factor, $h<1$ is the transmission-line attenuation coefficient, $\mathcal{R}$ is the responsivity of the square-law detector, with units of $\rm V/W$, $P_L$ is the local-oscillator power, and $Z_D$ is the impedance at the detection terminal.

Assuming that all noise sources are statistically independent, the total system noise can be expressed as the quadrature sum of the random noise, pump-source amplitude noise, pump-source phase noise and shot noise
\begin{align}
    \sigma_N=\sqrt{\sigma_{P_N}^2+\sigma_{A1}^2+\sigma_{P,\phi 1}^2+\sigma_{\rm shot1}^2},
\end{align}
where $\sigma_{A1}$
  and $\sigma_{P,\phi 1}$
  represent the root-mean-square (RMS) output noise values corresponding to the pump-source amplitude noise and phase noise, respectively, after double-balanced homodyne detection. Meanwhile, $\sigma_N$  and $\sigma_{\rm shot1}$ denote the RMS output noise values of the equivalent thermal noise from the cavity and transmission chain and the shot noise after double-balanced homodyne detection, respectively.  Based on the signal and noise propagation described above, the system noise can be approximately expressed as
\begin{align}
\sigma_N
\simeq&
\frac{8\mathcal{R}^2P_L}{Z_D}
\left(
\eta hP_{T,\rm cav}
+hP_{N,\rm ant}
+P_{N,\rm line}
\right)
\nonumber\\
&+
\frac{8\eta h\mathcal{R}^2P_L}{Z_D}
\left(
\sigma_A+\sigma_{P,\phi}+P_{\rm shot1}
\right).
\end{align}
Substituting the above expressions into the Dicke radiometer equation
\begin{align}
    {\rm SNR_{Dicke}}=\frac{\eta h(P_s
    ^{(1)}+P_s
    ^{(2)})\sqrt{t_{in}\Delta f}}{\eta hP_{T,cav}+hP_{N,ant}+P_{N,line}+\eta h(\sigma_A+\sigma_{P,\phi}+P_{\rm shot1})}
\end{align}
where $t_{in}$ and $\Delta f$ denote the total integration time and the bandwidth of a single measurement, respectively. For the parameter range considered in this work, we take $\tilde{B}_0=3{\rm nT},P_s^{(0)}\sim10^{-8}\rm W$, $\Delta f=100\rm Hz$, ${\rm RIN}=-120\rm dB/Hz$, and $S_\phi\sim10^{-10}\rm rad^2/Hz$, these parameters yield $\sigma_A\sim10^{-13}\rm W$ and $\sigma_{P,\phi}\sim10^{-12}\rm W$, respectively. In contrast, even at room temperature, the cavity thermal-noise power is only on the order of $10^{-18}\rm W$, which is several orders of magnitude smaller than the pump-source amplitude and phase noise. Therefore, for the parameter regime considered here, the system noise is dominated by the pump-source amplitude and phase noise, whereas the contributions of cavity thermal noise, antenna noise, and transmission-line thermal noise to the final SNR are negligible. Accordingly, the SNR reduces to
\begin{align}
     {\rm SNR_{Dicke}}=\frac{\sqrt{P_s^{(2)}t_{in}\Delta f}}{\sqrt{P_s^{(0)}}(\sqrt{{\rm RIN}\Delta f}+\sqrt{S_\phi \Delta f})+\sqrt{\hbar\omega/\tau}}.
\end{align}
When the system reaches the detection threshold, ${\rm SNR}_{\rm Dicke}\geq1$, the minimum detectable axion–photon coupling can be obtained as
\begin{align}\label{g1}
    g_{a\gamma\gamma}^{(1)}\ge\frac{\sqrt{P_s^{(0)}}(\sqrt{{\rm RIN}\Delta f}+\sqrt{S_\phi \Delta f})+\sqrt{\hbar\omega/\tau}}{\sqrt{\kappa t_{in}\Delta f}},
\end{align}
where $\kappa=P_{s}^{(2)}/g_{a\gamma\gamma}^2, ~P_{s}^{(2)}=g_{a\gamma\gamma}^2a_0^2\omega_aQ\bar{B}^2 C_1V/8\mu_0$. As the detection of axions at different frequencies requires resonant cavities of different volumes, the cavity volume is given by $V=9.556\times10^{24}/(f_{010}({\rm Hz}))^3({\rm m^3})$
   for a cavity length of $H=2R$.
  
For conventional Haloscope-types, which directly detect $P_{s}^{(2)}$, the SNR is mainly affected by shot noise and thermal noise and can be written as
\begin{align}
    {\rm SNR}=\frac{P_{s}^{(2)}\sqrt{t_{in}\Delta f}}{P_{\rm shot2}+P_{T,cav}},
\end{align}
where $P_{\text{shot2}}=(P_{s}^{(2)}\hbar\omega_a/\tau )^{1/2}$ is the shot noise power and $\tau\sim 10^{-4}$s denotes the detector response time.
 . 
Again, with $\rm SNR \ge 1$, the constraint on the axion coupling constant 
$g_{a\gamma\gamma}$
reads: 
\begin{align}\label{g2}
g_{a\gamma\gamma}^{(2)}\ge \frac{1}{2\sqrt{\kappa t_{in}\Delta f}}\left(\sqrt{\frac{\hbar\omega_a}{\tau}}+\sqrt{\frac{\hbar\omega_a}{\tau}+4 P_{T}}\right).
 \end{align}
\begin{figure}[htbp]
    \centering
    \begin{subfigure}[b]{0.23\textwidth}
        \centering
        \includegraphics[width=\linewidth]{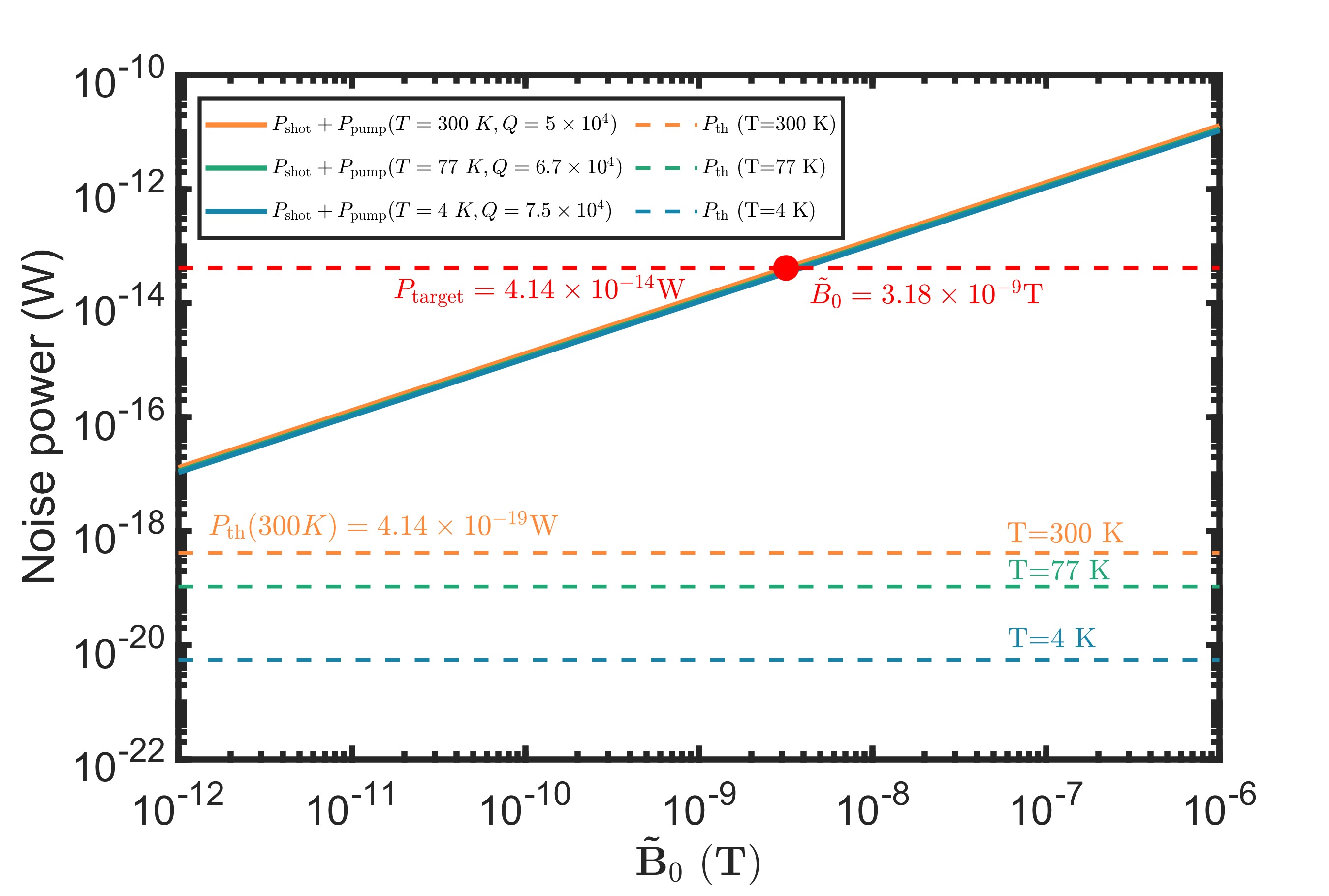}
        \caption{}
        \label{fig21}
    \end{subfigure}
    \hfill
    \begin{subfigure}[b]{0.24\textwidth}
        \centering
        \includegraphics[width=\linewidth]{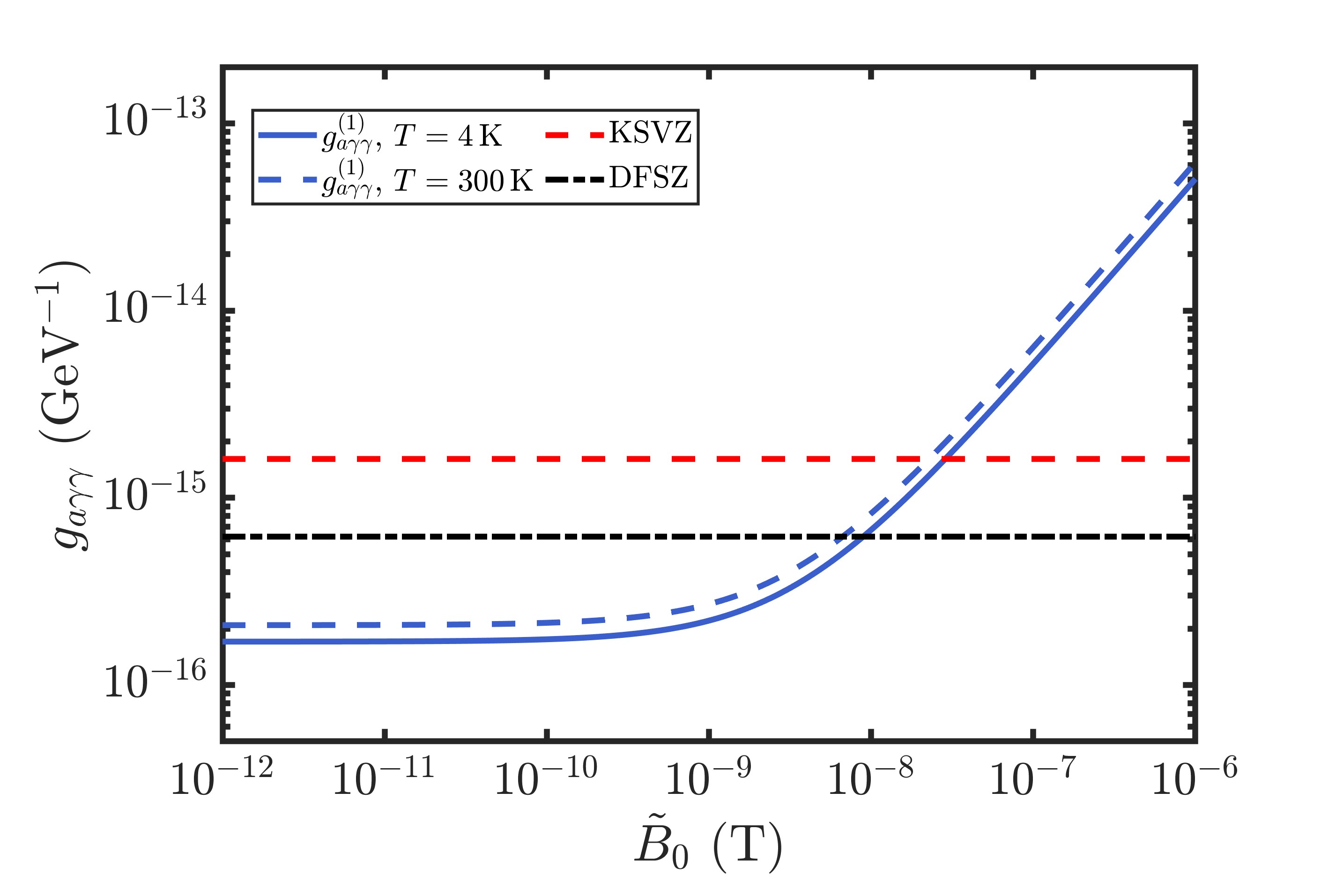}
        \caption{}
        \label{fig22}
    \end{subfigure}

    \caption{  
    Numerical analysis of the first-order power detection noise and the corresponding first-order sensitivity for different amplitudes of the seed field \(\tilde{B}_0\). (a) The seed-field-induced noise, shot noise, and system equivalent thermal noise at \(\tilde{B}_0=3.18\times10^{-9}{\rm T}\); the red dashed line denotes the target signal noise power \(P_{\rm target}=10^5P_{\rm th}(=300K)\), which exceeds the thermal noise power at 300 K. (b) For \(\tilde{B}_0<10^{-8}{\rm T}\), the first-order sensitivity \(g_{a\gamma\gamma}^{(1)}\) can exceed the DFSZ model prediction; further enhancement of the seed field will degrade this first-order sensitivity.}
    \label{fig2}
\end{figure}
Next, a central question arises: how much pump power is required for the pre-excited field to effectively suppress thermal noise? To answer this, we define a reference threshold \(P_{\rm target} \equiv 10^5P_{\rm th}(T=300{\rm K})\), i.e., five orders of magnitude above the thermal noise power at room temperature. This threshold ensures that, once the pump-induced noise $P_{\rm pump}=\sigma_A+\sigma_{P,\phi}$ and $P_{\rm shot1}$ exceeds it, the thermal noise becomes negligible in practice. Specifically, in Figure \ref{fig21} we show how this threshold can be surpassed for different values of \(\tilde{B}_{0}\). It is seen that, the crossover occurs at \(\tilde{B}_{0}=3.18\times10^{-9}\,{\rm T}\), above which the thermal noise is effectively suppressed. Furthermore, figure \ref{fig22} shows how the achievable sensitivity by the present first-order power detection can surpass the DFSZ prediction, when $\tilde{B}_0 < 10^{-8} \mathrm{T}$. Combining Figs.~\ref{fig21} and~\ref{fig22}, we find that the pump magnetic-field strength required to simultaneously satisfy the thermal-noise suppression condition and surpass the DFSZ prediction should lie in the range $3{\rm nT}<\tilde{B}_0<10 {\rm nT}$.
 In a typical cavity, the field amplitude relates to injected power by:$
\tilde{B}_{0} \sim \sqrt{{\mu_0 Q P_{\rm in}}/({\omega V})}$.
This indicates that only: the microwave source power
\(
P_{\rm in} \sim 2.6\times10^{-8}~{\rm W} \sim 26~{\rm nW}
\) is required for \(\tilde{B}_{0}=3\times 10^{-9}~{\rm T}\).  
At this optimal pre-excitation point, the equivalent noise temperature is:
\begin{align}
T_{\rm eff} = \frac{P_{\rm pump}}{k_B\Delta f} \sim 0.7~{\rm K},
\end{align}
the equivalent noise temperature is already below the 4 K physical temperature of conventional cryogenic haloscopes. This represents the quantum noise limit achievable with our scheme: it is lower than the level attained by conventional schemes at 4 K under thermal-noise limitation, but does not fall below the fundamental quantum lower bound (i.e., vacuum fluctuations $ \hbar\omega/2 $).

\begin{figure}[htbp]
\centering
\includegraphics[width=0.45\textwidth]{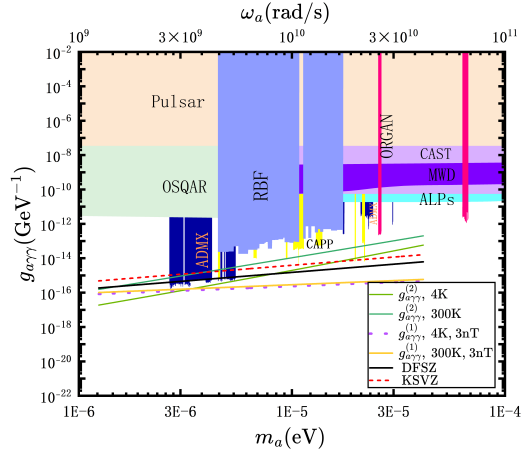}
\caption{Sensitivity comparison between the proposed scheme with seed field and the conventional Haloscopes without seed fields. The cavity quality factor \(Q\) is set to \(7.5\times10^4\) at \(T=4\) K and \(5\times10^4\) at \(T=300\) K. The results show that, even at room temperature (300 K), the achievable sensitivity of the present upgrade scheme (yellow line) can fall below the DFSZ prediction line over the 300 MHz–10 GHz range, indicating that room-temperature axion detection based on this upgrade scheme is feasible.}
\label{fig3}
\end{figure}

 Using conventional Haloscope parameters: \(\bar{B}=8\) T, \(\tau=10^{-4}\) s, \(\Delta f=100\) Hz~\cite{ADMX0}, \(t_{\rm in}=100\) s, and \(V=9.556\times10^{30}/[\omega_0/(2\pi)]^3~{\rm cm^3}\), we construct the axion sensitivity plot shown in Figure.~\ref{fig3}, incorporating the achievable sensitivity (Eq.~\ref{g1}) of the upgraded detector, the conventional second-order sensitivity (Eq.~\ref{g2}), the theoretical predictions of the KSVZ~\cite{KSVZ1,KSVZ2} and DFSZ~\cite{DFSZ1,DFSZ2} models, and the existing experimental bounds.
It is seen clearly that, 
the conventional sensitivity is strongly temperature-dependent, with lower cryogenic temperatures yielding more stringent limits. 
In contrast, for our proposed scheme, the noise induced by \(\tilde{B}_{0}\) overwhelmingly dominates over thermal noise, pushing the in-cavity experimental environment to the quantum noise limit. Consequently, even at room temperature (300 K) with \(\tilde{B}_{0}=3~{\rm nT}\), the upgraded sensitivity exceeds both the DFSZ and KSVZ model predictions across the entire frequency range from 300 MHz to 10 GHz.
Figure~\ref{fig3} clearly shows that the achievable sensitivity based on second-order power detection of the axion response signal is consistent with that expected from standard axion Haloscope experiments, exhibiting a strong temperature dependence: lower temperatures yield higher sensitivity. 

Consequently, in the low-frequency band (below 1.9 GHz), conventional Haloscope experiments require cryogenic environments (e.g., 100 mK for ADMX, 50 mK for CAPP) to exceed the KSVZ prediction line and reach the DFSZ line; in the high-frequency band (from approximately 1.9 GHz to 10 GHz), even operating at 4 K, they can only reach the KSVZ line but fall short of the DFSZ line.
In contrast, with the proposed upgrade scheme where the shot noise introduced by $\tilde{B}_{0}$ far exceeds the thermal noise, the quantum noise limit of axion detection sensitivity is pushed to a lower level. As a result, even at room temperature (300 K), the achievable sensitivity can simultaneously surpass both the KSVZ and DFSZ prediction lines across the entire frequency band.

\section{Conclusions and Discussions.}
We have achieved a paradigm shift in axion dark matter detection: from second-order power detection to first-order power detection, from passive waiting to active amplification. This paradigm shift does not modify the Primakoff mechanism of the axion electromagnetic response and thus the physical origin of axion--photon conversion; it coherently enhances the detectability of the same axion-induced response signal. Unlike external coherent amplification, the linear amplification in our scheme is achieved in situ inside the cavity, as the pre-excited field and the axion response share the same cavity mode and the same resonance frequency, ensuring strict frequency coherence without external locking. This is the physical prerequisite for the measurability of the first-order power signal.

Moreover, this paradigm shift is not a distant vision but a practical upgrade for existing experiments, as all the technologies are mature and all the components can be integrated into existing Haloscope platforms. The proposed scheme should therefore be understood as an in situ coherent enhancement of the detectability of the standard axion electromagnetic response, rather than a replacement or modification of the Primakoff mechanism.
This paradigm shift yields three distinct achievements.
First, the detectable power amplitude is enhanced by eight orders of magnitude relative to conventional Haloscopes. This addresses the question of whether the signal is measurable, bringing the axion response above the noise floor. Importantly, this enhancement is achieved by coherently amplifying the readout of the same Primakoff-induced response, not by altering the axion--photon conversion mechanism.
Second, the pre-excitation field reduces the system's equivalent quantum noise limit from room temperature (300 K) to 0.7 K, below the physical temperature of conventional schemes at 4~K. This is the physical origin of the sensitivity gain, as the first-order amplification provides a measurable signal and the lower quantum noise limit, arising in the seed-field noise-dominated regime, provides the higher achievable sensitivity.
Third, despite the random phase of the axion field (and hence that of its electromagnetic response signal), the amplitude of the first-order power signal can be precisely extracted via quadrature-balanced homodyne detection, because our scheme measures the amplitude rather than the phase. This stands in contrast to conventional external phase-sensitive amplification, where the random phase fluctuations of the signal would introduce significant uncertainties; a fundamental issue that phase-sensitive amplification alone cannot resolve, regardless of the available gain. Quadrature-balanced homodyne detection measures the length of the signal vector in the IQ plane, a quantity independent of the stochastic phase.
Collectively, these three achievements yield a substantial enhancement in detectability and a net sensitivity gain conferred by the lower quantum limit, promising for the first time to bring the achievable room-temperature sensitivity below the predictions of both the KSVZ and, remarkably, the more stringent DFSZ model lines. This marks the transition of axion dark matter detection from chasing theory to surpassing theory.

This four-decade-long quest for axions may well be standing at a historic turning point: from "building colder cryostats and waiting for longer integration times" to "leveraging mature room-temperature technology to actively amplify the response signal, thereby achieving higher sensitivity at a lower equivalent quantum noise limit." Our scheme provides a drop-in upgrade path for every existing Haloscope experiment. It should be clearly stated that the proposed approach does not change the Primakoff mechanism or the physical origin of axion--photon conversion; it is an in situ coherent enhancement of the detectability of that response, with the sensitivity gain arising from the seed-field shot-noise-dominated regime. The remaining question, perhaps, is no longer "whether we can achieve it," but "when we choose to begin". The answer to that question depends on whether we are willing to incorporate this mature scheme into the upgrade road-map of next-generation Haloscope experiments.

\section*{Acknowledgements}
This work was partially supported in part by Central Guidance Funds for Local Science and Technology Development No. 2025ZYD0154, the National Key Research and Development Program of China under Grant No.~2021YFA0718803, the National Natural Science Foundation of China under Grant 
No.~P110325G02011, No.~12505083, the Fundamental Research Funds for the Central Universities under Grant No.~2682024CX048, and the Natural Science Foundation of Sichuan under Grant No.~2025ZNSFSC0857. The National Natural Science Foundation of China (Grant No. 12575030).

\appendix

\section{Derivation of Zeroth- and First-Order Axion's Electromagnetic Responses}\label{Appendix A}
When the axion field is absent, the electromagnetic field in a microwave cavity (the zeroth‑order axion response) obeys the conventional Maxwell equations
\begin{equation}
    \left\{
    \begin{aligned}
  &\nabla\times\bm{B}^{(0)}+\frac{1}{c^2}\partial_t\bm{E}^{(0)}=\mu_0\bm{J},\\
  &\nabla\cdot\bm{E}^{(0)}=\frac{\rho}{\varepsilon_0},
  \end{aligned}
    \right.
\end{equation}
where $\bm{J}(\bm{x}, t) =J_0
\bm{J}(\bm{x})\cos{(\omega_0t)}$ is the external excitation current source, and $\rho$ is the free charge inside the cavity. Tuning the frequency $\omega_0$  of the excitation source $\bm{J}(\bm{x},t)$ to resonate with the $\mathrm{TM}_{010}$ mode frequency $\omega_m$ and solving the zeroth-order wave equation yields the excitation of the target cavity mode by the source, producing the zeroth-order magnetic field $\tilde{B}_0(\bm{x}, t)$. The time-dependent part of the steady-state excited field for the $\mathrm{TM}_{010}$ mode inside the cavity satisfies the driven-dissipation equation
\begin{align}\label{A2}
    &(\frac{d^2}{dt^2}+\gamma_m\frac{d}{dt}+\omega_m^2)\tilde{B}_0(t)=c^2\mu_0C_{0}J_0\bm{X}_m(\bm{x})\cos{(\omega_0t)},
\end{align}
where $C_{0}=\int_V\bm{X}_m^*(\bm{x})\cdot\nabla\times\bm{J}(\bm{x})dV$ is the overlap coefficient between the excitation source and the normalized spatial part $\bm{X}_m(\bm{x})$ of the $\mathrm{TM}_{010}$ mode, $\gamma_m=\omega_{m}/Q$ is the dissipation coefficient, and $\omega_m$ is mode frequency of the $\mathrm{TM}_{010}$ mode inside a vacuum cylindrical resonant cavity, and $Q$ is the quality factor of the cavity for this mode. The steady-state solution of Eq.~\eqref{A2} is
\begin{align}
\tilde{B}_0(\bm{x},t)=\frac{c^2\mu_0J_0C_0k_c\bm{X}_m(\bm{x})}{\omega_m^2-\omega_0^2-i\omega_0\gamma_m}\cos{(\omega_0t)},
\end{align}
where $k_c=x_0^1/R=21$ $\text{m}^{-1}$, $x_0^1$ is the first root of the zeroth-order Bessel function, and $R=0.11459$ \text{m} is the radius of the resonant cavity. Using the electromagnetic field relations, the zeroth-order axion response electric field intensity inside the cavity is obtained as
\begin{align}\label{A4}
   \tilde{E}_0(\bm{x},t)=\frac{c^3\mu_0J_0C_0k_c\bm{X}_m(\bm{x})}{\omega_m^2-\omega_0^2-i\omega_0\gamma_m}\sin{(\omega_0t)}\bm{e}_z.
\end{align}
The zeroth-order axion response magnetic field inside the cavity is:
\begin{equation}\label{A5}
    \bm{B}^{(0)}(\bm{x},t)=\tilde{B}_0(\bm{x},t)\bm{e}_\theta+\bar{B}(\bm{x})\bm{e}_z,
\end{equation}
where $\bar{B}$ is a steady strong magnetic field along the axial direction $\bm{e}_z$ of the cylindrical resonant cavity.

As the axion field enters the microwave cavity, axion‑photon energy conversion proceeds via the inverse Primakoff effect. The effective current induced by the axion field excites the target‑mode electromagnetic field within the cavity, which is governed by the axion‑modified first‑order Maxwell equations:
\begin{equation}
    \left\{
    \begin{aligned}
&\nabla\times\bm{B}^{(1)}+\frac{1}{c^2}\partial_t\bm{E}^{(1)}=-\frac{1}{4c}g_{a\gamma\gamma}\bm{B}^{(0)}\partial_{t}a,\\
  &\nabla\cdot\boldsymbol{E}^{(1)}=\frac{1}{4}cg_{a\gamma\gamma}\bm{B}^{(0)}\cdot\nabla a.
  \end{aligned}
    \right.
\end{equation}
Given that the axion field is approximately isotropic, implying $\nabla a\sim 0$, the above equation simplifies to:
\begin{equation}
\left\{
\begin{aligned}
&\nabla\times\boldsymbol{B}^{(1)}+\frac{1}{c^2}\partial_t\boldsymbol{E}^{(1)}=-\frac{1}{4c}g_{a\gamma\gamma}\bm{B}^{(0)}\partial_ta,\\
&\nabla\cdot\boldsymbol{E}^{(1)}=0.
\end{aligned}
\right.
\end{equation}
Under the Lorentz gauge, the electromagnetic potentials of the first-order axion electromagnetic response field satisfy the equation:
\begin{align}    
&\nabla^2\bm{A}^{(1)}-\frac{1}{c^2}[\partial_t^2\bm{A}^{(1)}- \gamma \partial_t\bm{A}^{(1)}]=\nonumber\\    
&\quad -\frac{g_{a\gamma\gamma}}{4c}[(\bar{B}\partial_ta)\bm{e}_z+(\tilde{B}_0\partial_ta)\bm{e}_\theta],
\end{align}
where the dissipation coefficient $\gamma=\omega/Q$ is related to the excited cavity mode frequency $\omega$ and the cavity quality factor $Q$. The first-order axion-response electromagnetic field is obtained by solving the axion-modified first-order Maxwell equations. The axion field can be written in general as: $a(\bm{x},t)=\mathrm{Re}[a_0\exp(i\bm{k}_a\cdot\bm{x}-i\omega_a t-i\phi_a)]\sim a_0\cos{(\omega_a t+\phi_a)}$, where $a_0=(2\rho_a\hbar c^5/\omega_a^2)^{1/2}$ is the axion field amplitude, $\omega_a$ and $\bm{k}_a$ are its frequency and wave vector, $\rho_a=0.5\times10^{-24}~\mathrm{g\cdot cm^{-3}}$ is the dark matter axion density, and $\phi_a$ is the axion phase (which randomly varies in $[0,2\pi]$ for times longer than the axion coherence time $\tau_a\sim 2\pi \hbar c^2/(m_a c^2\sigma_v^2)$ with $~\sigma_v=10^{-3}c$).

Using the zeroth-order axion response field in Eq.~\eqref{A5}, the effective axion currents in the $\bm{e}_\theta$ and $\bm{e}_z$ directions can be separately formulated, and then the first-order response electromagnetic field intensities are derived. First, the first-order magnetic vector potential excited by the $\bm{e}_z$-direction effective current can be written as:
\begin{align}
   &A_z^{(1)}(\bm{x},t)= - a_0g_{a\gamma\gamma} C_1\bm{X}_{m}(\bm{x})  \frac{c\omega_a\bar{B}}{4D_m}\nonumber\\
        &\quad\times\left[\Delta_m\sin{(\omega_at+\phi_a)}-\Gamma_m\cos{(\omega_at+\phi_a)}\right],
\end{align}
where $D_m=\Delta_m^2+\Gamma_m^2$, $\Delta_m=\omega_m^2-\omega_a^2$, $\Gamma_m=\gamma_m\omega_a$, and 
$C_1=\int_V \bm{X}_m^*(\bm{x})\cdot \bm{\bar{B}}(\bm{x}) dV$ is the overlap integral between the effective current (induced by the axion under a strong magnetic field) and the $\mathrm{TM}_{010}$ mode. 
This result is consistent with the axion electromagnetic response in a conventional Haloscope-type experiment. 
Consequently, using the relations between the vector potential $\bm{A}^{(1)}$ and the fields $\bm{E}^{(1)}$, $\bm{B}^{(1)}$, the first-order electromagnetic field strengths can be expressed as follows:
\begin{align}\label{A10}
   & E_{mz}^{(1)}(\bm{x},t)
    = a_0g_{a\gamma\gamma} C_1\bm{X}_{m}(\bm{x})  \frac{c\omega_a^{2}\bar{B}}{4D_m}\bm{e}_z\nonumber\\
        &\quad\times\left[\Gamma_m\sin{(\omega_at+\phi_a)}+\Delta_m\cos{(\omega_at+\phi_a)}\right],
\end{align}
and
\begin{align}
    &B^{(1)}_{m\theta }(\bm{x},t)
    = a_0g_{a\gamma\gamma}  C_1k_c\bm{X}_m(\bm{x})\frac{c\omega_a\bar{B}}{4D_m} \bm{e}_\theta\nonumber\\
        &\quad\times\left[\Delta_m\sin{(\omega_at+\phi_a)}-\Gamma_m\cos{(\omega_at+\phi_a)}\right],
\end{align}
where $k_c=x_0^1/R$. Secondly, in the $\bm{e}_\theta$ direction, the first-order magnetic vector potential excited by the effective current induced by the axion response can be solved as:
\begin{align}
     &A_{\theta}^{(1)}(\bm{x},t)=-a_0g_{a\gamma\gamma}C_2\bm{X}_    l(\bm{x})\frac{c\omega_a\tilde{B}_0}{8D_l}\nonumber\\
&\quad\times\left[ \Delta_l\cos{(\omega_+t+\phi_a)}
    +\Gamma_l\sin{(\omega_+t+\phi_a)} \right],
\end{align}
where $D_l=\Delta_l^2+\Gamma_l^2$, $\Delta_l=\omega_l^2-\omega_+^2$, $\Gamma_l=\gamma_l\omega_+$, and the sum-frequency term $\omega_+=\omega_a+\omega_0$ can excite other modes in the cavity, such as the $\text{TE}_{011}$ mode. Thus, $C_2=\int_V \bm{X}_{l}^{*}(\bm{x})\cdot \bm{X}_m(\bm{x}) dV$ is the overlap integral between the effective current induced by $\tilde{B}_0$ and the $\text{TE}_{011}$ mode.
Meanwhile, $\bm{X}_l(\bm{x})$, $\omega_l$ and $\gamma_l=\omega_l/Q$ represent the spatial distribution function, mode frequency, and dissipation coefficient of the $\text{TE}_{011}$ mode, respectively. Therefore, after the axion interacts with the steady-state excitation magnetic field 
 $\tilde{B}_0$, the first-order response electromagnetic field is given by:
\begin{align}
    &E_{l\theta }^{(1)}(\bm{x},t)
=a_0g_{a\gamma\gamma}  C_2\bm{X}_{l}(\bm{x})\frac{c\omega_a\tilde{B}_0\omega_+}{8D_l}\bm{e}_\theta\nonumber\\
&\quad\times\left[ \Gamma_l\cos{(\omega_+t+\phi_a)}-\Delta_l\sin{(\omega_+t+\phi_a)}
     \right].
\end{align}
and
\begin{equation}
    \left\{
\begin{aligned}
    &B_{l\rho }^{(1)}(\bm{x},t)=a_0g_{a\gamma\gamma}C_2\frac{c\omega_a\tilde{B}_0}{8D_l}\frac{\pi}{H}\bm{X}_l(\bm{x})\bm{e}_\rho\\
&\quad\times\left[ \Delta_l\cos{(\omega_+t+\phi_a)}+\Gamma_l\sin{(\omega_+t+\phi_a)} \right],\\
    &B_{lz}^{(1)}(\bm{x},t)=-a_0g_{a\gamma\gamma}k_cC_2\frac{c\omega_a\tilde{B}_0}{8D_l} \bm{X}_l(\bm{x})\bm{e}_z\\
&\quad\times\left[ \Delta_l\cos{(\omega_+t+\phi_a)}    +\Gamma_l\sin{(\omega_+t+\phi_a)} \right],
\end{aligned}
\right.
\end{equation}
where $H=0.12443\rm m$ is the axial length of the microwave cavity. Thus, the first-order axion-induced electromagnetic response fields inside the cavity can be generally expressed as:
\begin{equation}\label{A15}
    \left\{
\begin{aligned}
   & \bm{E}^{(1)}(\bm{x},t)=E_{l\theta}^{(1)}\bm{e}_\theta+E_{mz}^{(1)}\bm{e}_z,\\\
    &\bm{B}^{(1)}(\bm{x},t)=B_{l\rho}^{(1)}\bm{e}_\rho+B_{m\theta}^{(1)}\bm{e}_\theta+B_{lz}^{(1)}\bm{e}_z.
\end{aligned}
    \right.
\end{equation}
Consequently, using Eq.~\eqref{A5}, the total electromagnetic field within the cavity can be written as:
\begin{equation}
    \left\{
\begin{aligned}
    \bm{E}(\bm{x},t)=\bm{E}^{(0)}(g_{a\gamma\gamma}^0)+\bm{E}^{(1)}(g_{a\gamma\gamma})+\mathcal{O}(g_{a\gamma\gamma}^2),\\[3pt]
    \bm{B}(\bm{x},t)=\bm{B}^{(0)}(g_{a\gamma\gamma}^0)+\bm{B}^{(1)}(g_{a\gamma\gamma})+\mathcal{O}(g_{a\gamma\gamma}^2).
\end{aligned}
\right.
\end{equation}
This corresponds to the electromagnetic field strengths of the zero- and first-order axion electromagnetic responses inside the cavity.

\section{The Power of the Axion's Electromagnetic Response Signal}\label{Appendix B}
From the zero- Eq.~\eqref{A5} and first-order Eq.~\eqref{A15} axion electromagnetic response fields, the signal power expanded in the axion coupling constant $g_{a\gamma\gamma}$ is directly given by:
\begin{align}
   P_s(t)&=\frac{\gamma}{2}
\int_V(\varepsilon_0\bm{E}\cdot\bm{E}+\frac{1}{\mu_0}\bm{B}\cdot\bm{B})dV=\sum_{i=0}^2P_s^{(i)}(t)(g_{a\gamma\gamma}^{(i)},t),
\end{align}
where
$
P_s^{(0)}(t)=\gamma_m
\int_V(\varepsilon_0\tilde{E}_0^2+\tilde{B}_0^2/\mu_0)dV/2$ is the zero-order power. This term contains no axion information and is entirely determined by the zero-order response electromagnetic field strength, thus representing a relatively large background power.
\begin{align}
&P_s^{(1)}(t)=\frac{\gamma}{2}\int_V\left[\varepsilon_0\tilde{E}_0E_{mz}^{(1)}+\frac{1}{\mu_0}(\tilde{B}_0B_{m\theta}^{(1)}+\bar{B}B_{lz}^{(1)})\right]dV\nonumber\\  &=\int_V\frac{\gamma_m\varepsilon_0c^2\omega_a^2}{8D_m}a_0g_{a\gamma\gamma}\tilde{B}_0\bar{B}\bm{X}_m^2(\bm{x})C_1dV\nonumber\\
    &\quad\times\sin{(\omega_0t)}\left[\Gamma_m\sin{(\omega_at+\phi_a)}+\Delta_m\cos{(\omega_at+\phi_a)}\right]\nonumber\\
    &+\int_V\frac{\gamma_mc\omega_ak_c}{8\mu_0D_m}a_0g_{a\gamma\gamma}\tilde{B}_0\bar{B}\bm{X}_m^2(\bm{x})C_1dV\nonumber\\
    &\quad\times\cos{(\omega_0t)}\left[\Delta_m\sin{(\omega_at+\phi_a)}-\Gamma_m\cos{(\omega_at+\phi_a)}\right]\nonumber\\
    &+\int_V\frac{\gamma_lc\omega_ak_c}{16\mu_0D_l}a_0g_{a\gamma\gamma}\bar{B}\tilde{B}_0\bm{X}_l(\bm{x})C_2dV\nonumber\\
    &\quad\times\left[\Delta_l\cos{(\omega_+t+\phi_a)}+\Gamma_l\sin{(\omega_+t+\phi_a)}\right],
\end{align}
the first-order instantaneous power is proportional to the sensitivity $g_{a\gamma\gamma}$. At resonance, we have $\omega_0=\omega_a=\omega_m$, $D_m=\Gamma_m^2,\Delta_m=0$, and $\omega_+=\omega_l$, $D_l=\Gamma_l^2,\Delta_l=0$. Consequently,
\begin{align}\label{B3}
&P_s^{(1)}(t)=[\int_V\frac{\gamma_m\omega_a^2}{8\mu_0\Gamma_m}a_0g_{a\gamma\gamma}|\tilde{B}_0||\bar{B}|\bm{X}_m^2(\bm{x})C_1dV\nonumber\\
&+\int_V\frac{\gamma_l\omega_a^2}{16\mu_0\Gamma_l}a_0g_{a\gamma\gamma}|\bar{B}||\tilde{B}_0|\bm{X}_l(\bm{x})C_2dV]\cos{(2\omega_at+\phi_a)},
\end{align}
with the amplitude:
\begin{align}\label{B4}
P_{s}^{(1)}=\frac{\omega_a}{8\mu_0}a_0g_{a\gamma\gamma}|\tilde{B}_0||\bar{B}|V(C_1+\frac{C_2}{4})\propto |\tilde{E}_0||E_{mz}^{(1)}|,
\end{align}
which is detectable (see the \ref{Appendix C} below). The following parameters are adopted: 
$\omega_a=2\pi\times1$ \text{GHz}, $Q=10^4$, $|\bar{B}|=8$ \text{T}, $\tilde{B}_0=3~\mu \text{T}$, $C_0=C_1=C_2=1$ and $V=\pi R^2H=5.13$ \text{T}. Consequently, Eq.~\eqref{B4} evaluates to:
\begin{align}
&P_{s}^{(1)}=3.24\times10^{-17}\mathrm{W}\left(\frac{g_\gamma}{0.36}\right)\left(\frac{10^{12}\mathrm{GeV}}{f_a}\right)\left(\frac{\omega_a}{2\pi~\mathrm{GHz}}\right)\nonumber\\
&\times\left(\frac{\rho_a}{\frac{1}{2}\times10^{-21}\mathrm{(\frac{kg}{m^3})}}\right)\left(\frac{\tilde{B}_0}{3\mathrm{n T}}\right)\left(\frac{\bar{B}}{8\mathrm{T}}\right)\left(\frac{Q}{10^4}\right)\left(\frac{V}{5.13\mathrm{L}}\right).
\end{align}
The second-order power containing axion information is:
\begin{align}\label{B6}
    P_s^{(2)}(t)&=\frac{\gamma}{2}\int_V[\varepsilon_0(|E_{mz}^{(1)}|^2+|E_{l\theta}^{(1)}|^2)\nonumber\\
    &\quad\quad\quad+\frac{1}{\mu_0}(|B_{l\rho}^{(1)}|^2+|B_{m\theta}^{(1)}|^2+|B_{lz}^{(1)}|^2)]dV\nonumber\\
    &=P_m\left[\Gamma_m\sin{(\omega_at+\phi_a)}+\Delta_m\cos{(\omega_at+\phi_a)}\right]^2\nonumber\\
    &\quad+P_l\left[\Gamma_l\cos{(\omega_+t+\phi_a)}-\Delta_l\cos{(\omega_at+\phi_a)}\right]^2\nonumber\\
    &\quad+P_l\left[\Delta_l\cos{(\omega_+t+\phi_a)}+\Gamma_l\sin{(\omega_+t+\phi_a)}\right]^2\nonumber\\
    &\quad+P_m\left[\Delta_m\sin{(\omega_at+\phi_a)}-\Gamma_m\cos{(\omega_at+\phi_a)}\right]^2,
\end{align}
where,
$P_m=\int_V\gamma_m\varepsilon_0(a_0g_{a\gamma\gamma}c|\bar{B}|\omega_a^2\bm{X}_m(\bm{x}))^2C_1dV/{8D_m^2}$ and $P_l=\int_V\gamma_l\varepsilon_0(a_0g_{a\gamma\gamma}c|\tilde{B}_0|\omega_a\omega_+\bm{X}_l(\bm{x}))^2C_2dV/{16D_l^2}$.
At resonance, $\omega_0=\omega_a=\omega_m$ and $\omega_+=\omega_l$, the second-order response power of the axion is:
\begin{align}\label{B7}
    P_s^{(2)}    &=\frac{\gamma_m}{8\mu_0\Gamma_m^2}(a_0g_{a\gamma\gamma}|\bar{B}|\omega_a^2)^2C_1V+\frac{\gamma_l}{16\mu_0\Gamma_l^2}(a_0g_{a\gamma\gamma}|\tilde{B}_0|\omega_a\omega_+)^2C_2V\nonumber\\
    & \simeq\frac{1}{8\mu_0}a_0^2g_{a\gamma\gamma}^2\bar{B}^2\omega_aC_lQV.
\end{align}
Given that $\tilde{B}_0\ll\bar{B}$, the second term in Eq.~\eqref{B7} makes a negligible contribution. Under the same parameters, the second-order axion response power can be computed as:
\begin{align}
&P_s^{(2)}=3.63\times10^{-25}\mathrm{W}\left(\frac{g_\gamma}{0.36}\right)^2\left(\frac{10^{12}\mathrm{GeV}}{f_a}\right)^2\left(\frac{\bar{B}}{8\mathrm{T}}\right)^2\nonumber\\
&\times\left(\frac{2\rho_a}{10^{-21}kg\cdot m^{-3}}\right)\left(\frac{\omega_a}{2\pi~\mathrm{GHz}}\right)\left(\frac{Q}{10^4}\right)\left(\frac{V}{5.13\mathrm{L}}\right).
\end{align}
According to Eqs.~(\ref{B3},\ref{B4},\ref{B6},\ref{B7}), if there is no excitation of the $\mathrm{TM}_{010}$ mode magnetic field in the resonant cavity ($\tilde{B}_0=0$), then the first-order axion response power $P_{s}^{(1)}$
  becomes zero, while the second-order power $P_{s}^{(2)}$
  reverts to the conventional Haloscope-type.
  
\section{Quadrature-Balanced Homodyne Detection scheme for extracting the amplitude of the first-Order power signal $P_s^{(1)}(\phi_a,t)$.}\label{Appendix C}

\begin{figure}
    \centering
    \includegraphics[width=0.85\linewidth]{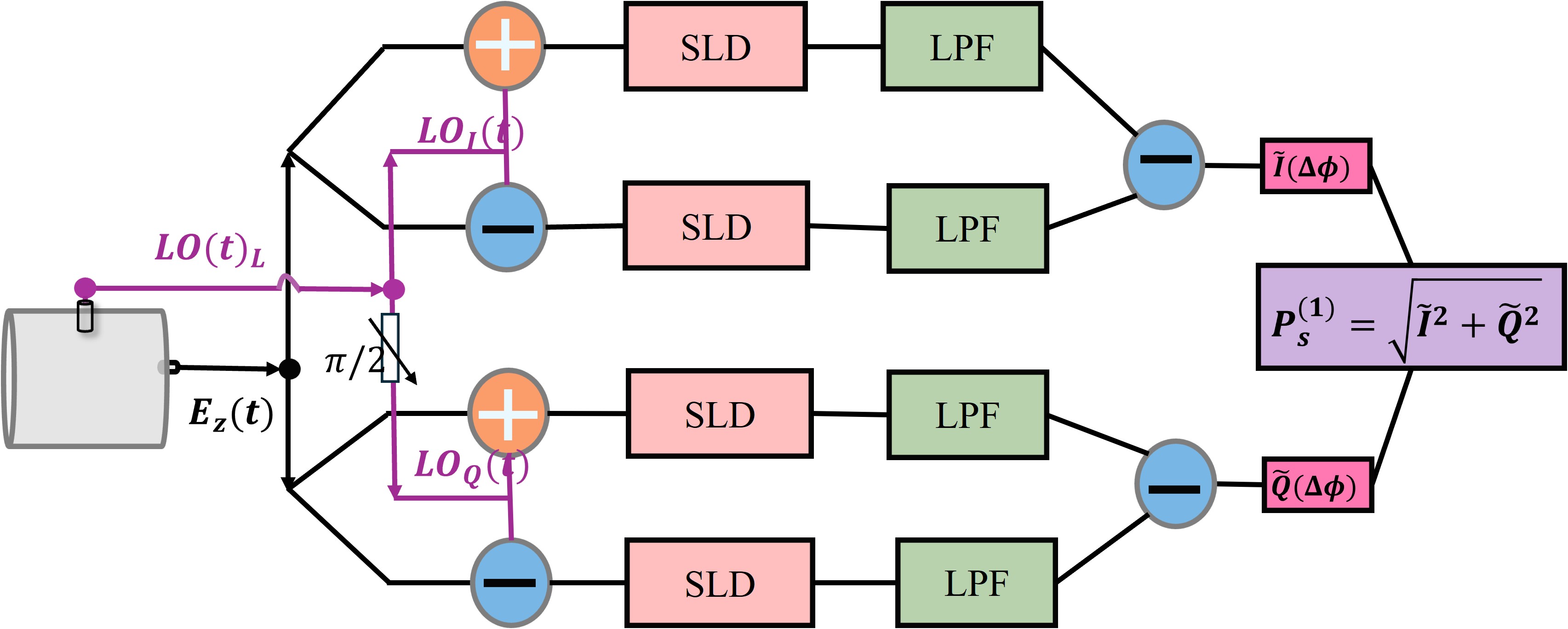}
    \caption{The QBHDS (Quadrature Balanced Homodyne Detection System). The plus denotes the signal sum channel, and the minus denotes the signal difference channel. SLD stands for square-law detector, and LPF stands for low-pass filter.}
    \label{fig5}
\end{figure}

From Eqs.~\eqref{A4} and \eqref{A10}, at the resonance condition of the axion field, drive, and $\mathrm{TM}_{010}$ mode frequencies, i.e., $\omega_0=\omega_a=\omega_m$ the total electric field strength along $\bm{e}_z$ is given by:
\begin{align}
E_z(t)&=\tilde{E}_0\sin{(\omega_0t+\phi_0)}+E_{mz}^{(1)}\sin{(\omega_0t+\phi_a)},
\end{align}
Consequently, the total power of the signal can be expressed as:
\begin{align}
    P_z&\propto \frac{1}{2}[(\tilde{E}_0)^2+(E_{mz}^{(1)})^2]+\tilde{E}_0E_{mz}^{(1)}\cos{(\Delta\phi)}\nonumber\\
&=P^{(0)}_s+P_s^{(2)}+P_s^{(1)},
\end{align}
where $\Delta\phi = \phi_a - \phi_0$ is the phase difference between the axion field phase $\phi_a$ and the seed field phase $\phi_0$, and in this article $\phi_0 = 0$. $P^{(0)}$
  is the power of the zero--order axion response signal, which is entirely determined by the pre‑excited steady‑state zero‑order field (independent of the axion); $P_s^{(2)}\propto g^2_{a\gamma\gamma}$ is the second‑order response power of the signal (i.e., the conventional Haloscope‑type detection power). Since the first-order axion electromagnetic response signal power $P_s^{(1)}\propto g_{a\gamma\gamma}$ (see Eq.~\eqref{B3} contains a random phase $\phi_a$, its statistical average vanishes. Consequently, it cannot be detected using conventional incoherent detection methods. To detect the amplitude $P_s^{(1)}$ of the axion first-order electromagnetic response power signal with random phase, which is amplified in situ by the steady-state field of the pre‑excited target mode (besides the externally amplified by using a low-noise quantum amplifier), we consider the quadrature balanced homodyne detection measurement system shown in Fig.~\ref{fig5}.
Among them
\begin{equation}
    \left\{
\begin{aligned}
    {\rm LO}_I(t)&=\tilde{E}_0\sin{(\omega_0t+\phi_0)},\\
     {\rm LO}_Q(t)&=\tilde{E}_0\cos{(\omega_0t+\phi_0)}.
\end{aligned}
\right.
\end{equation}
The two local oscillators operate at the same frequency as the zero-order electric field. The cavity signal of interest $E_z(t)$ is output via the receiving antenna and then split into two paths. The I‑path and Q‑path signals are respectively added to and subtracted from ${\rm LO}_{I}(t)$ and ${\rm LO}_{Q}(t)$. For example, for the I‑path signal, we obtain:
\begin{align}
    &V_{I+}(t)=E_z(t)+{\rm LO}_I(t)\nonumber\\
    &\quad=2\tilde{E}_0\sin{(\omega_0t+\phi_0)}+E_{mz}^{(1)}\sin{(\omega_0t+\phi_a)},\\
    &V_{I-}(t)=E_z(t)-{\rm LO}_I(t)=E_{mz}^{(1)}\sin{(\omega_0t+\phi_a)}.
\end{align}
They are respectively sent into the square-law detectors for squaring. Let the signals pass through the low-pass filters, we have:
\begin{equation}
    \left\{
\begin{aligned}
    P_{I+}=\langle V_{I+}^2\rangle_{\rm LPF}&=2(\tilde{E}_0)^2+\frac{(E_{mz}^{(1)})^2}{2}\\
    &\quad+2\tilde{E}_0E_{mz}^{(1)}\cos{(\Delta \phi)},\\
    P_{I-}=\langle V_{I-}^2\rangle_{\rm LPF}&=\frac{(E_{mz}^{(1)})^2}{2}.
\end{aligned}
\right.
\end{equation}
After differencing, it reads:
\begin{align}
    S_I=P_{I+}-P_{I-}=2(\tilde{E}_0)^2+2\tilde{E}_0E_{mz}^{(1)}\cos{(\Delta \phi)}.
\end{align}
Having accordingly subtracted the DC component, the term concerning the first-order power amplitude is derived as
\begin{align}
    \tilde{I}(\Delta\phi)=2\tilde{E}_0E_{mz}^{(1)}\cos{(\Delta \phi)}.
\end{align}
Correspondingly, for the Q‑branch signal, having respectively summed $E_z(t)$ with and subtracted ${\rm LO}_Q(t)$ from it, we subsequently obtain:
\begin{align}
    V_{Q+}(t)&=E_z(t)+{\rm LO}_Q(t)\nonumber\\
    &=\tilde{E}_0\sin{(\omega_0t+\phi_0)}+E_{mz}^{(1)}\sin{(\omega_0t+\phi_a)}\nonumber\\
    &\quad+\tilde{E}_0\cos{(\omega_0t+\phi_0)},\\
    V_{Q-}(t)&=E_z(t)-{\rm LO}_Q(t)\nonumber\\
&=\tilde{E}_0\sin{(\omega_0t+\phi_0)}+E_{mz}^{(1)}\sin{(\omega_0t+\phi_a)}\nonumber\\
    &\quad-\tilde{E}_0\cos{(\omega_0t+\phi_0)},
\end{align}
Making use of the auxiliary angle formula in trigonometry $\tilde{E}_0\sin{(\omega_0t+\phi_0)}\pm \tilde{E}_0\cos{(\omega_0t+\phi_0)}=\sqrt{2}\tilde{E}_0\cos{(\omega_0t+\phi_0\mp\frac{\pi}{4})}$, it can accordingly be simplified to:
\begin{equation}
    \left\{
\begin{aligned}
    V_{Q+}(t)&=\sqrt{2}\tilde{E}_0\sin{(\omega_0t+\phi_0-\frac{\pi}{4})}\\&+E_{mz}^{(1)}\sin{(\omega_0t+\phi_a)},\\
    V_{Q-}(t)&=\sqrt{2}\tilde{E}_0\sin{(\omega_0t+\phi_0+\frac{\pi}{4})}\\&+E_{mz}^{(1)}\sin{(\omega_0t+\phi_a)},
\end{aligned}
\right.
\end{equation}
After undergoing square‑law detection and low‑pass filtering, the powers of the signals read
\begin{equation}
    \left\{
\begin{aligned}
    P_{Q+}=\langle V_{Q+}^2\rangle_{\rm LPF}&=(\tilde{E}_0)^2+\frac{(E_{mz}^{(1)})^2}{2}\\
    &\quad+\sqrt{2}\tilde{E}_0E_{mz}^{(1)}\cos{(\Delta \phi+\frac{\pi}{4})},\\
    P_{Q-}=\langle V_{Q-}^2\rangle_{\rm LPF}&=(\tilde{E}_0)^2+\frac{(E_{mz}^{(1)})^2}{2}\\
    &\quad+\sqrt{2}\tilde{E}_0E_{mz}^{(1)}\cos{(\Delta \phi-\frac{\pi}{4})},
\end{aligned}
\right.
\end{equation}
Similarly, after differencing, the Q‑branch output pertaining to the first‑order signal power amplitude is readily obtained as:
\begin{align}
    S_Q&=P_{Q+}-P_{Q-}\nonumber\\&=\sqrt{2}\tilde{E}_0E_{mz}^{(1)}[\cos{(\Delta \phi+\frac{\pi}{4})}-\cos{(\Delta \phi-\frac{\pi}{4})}]\nonumber\\
    &=-2\tilde{E}_0E_{mz}^{(1)}\sin{(\Delta \phi)},
\end{align}
Namely, the power signal at the Q‑branch output is thereby expressed as
\begin{align}
    \tilde{Q}(\Delta\phi)=2\tilde{E}_0E_{mz}^{(1)}\sin{(\Delta \phi)}.
\end{align}
Combining $\tilde{I}(\Delta\phi)$
  and $\tilde{Q}(\Delta\phi)$
 , the amplitude $P_s^{(1)}$ in equation Eq.~\eqref{10} of the main text can accordingly be measured as:
\begin{align}
   P_s^{(1)}\propto P_{\rm out}=\sqrt{\tilde{I}^2(\Delta\phi)+\tilde{Q}^2(\Delta\phi)}=2\tilde{E}_0E_{mz}^{(1)}.
\end{align}
This indicates that, despite the randomness of the first--order power signal of the axion electromagnetic response, which consequently precludes its detectability, its amplitude can nevertheless be demodulated using the quadrature balanced homodyne detection scheme proposed herein. Consequently, its amplitude can be directly employed to evaluate the attainable detection sensitivity of the axion electromagnetic response signal, thereby effectively circumventing the difficulty encountered by conventional Haloscope‑type detectors in measuring the exceedingly weak second‑order power signal.

Naturally, in the foregoing detectability analysis, we have completely neglected the noise within the quadrature balanced homodyne measurement setup purely for brevity. Its influence is subsequently incorporated as thermal noise into the evaluation of the attainable detection sensitivity in the main text.

\section{Noise Analysis of a Double-Balanced Homodyne Axion Haloscope type}\label{appendix 4}
In this section, we mainly discuss the noise effects on the axion signal from its generation inside the cavity to the double-balanced homodyne detection stage, as well as the definition of the signal-to-noise ratio (SNR).

First, the readable signal power in the cavity can be expanded in powers of the axion coupling constant $g_{a\gamma\gamma}$ as follows:
\begin{align}
     P_s(t)=\sum _{i=0}^{2}P_s^{(i)}(t)=P_s^{(0)}(g_{a\gamma\gamma}^0,t)+P_s^{(1)}(g_{a\gamma\gamma},t)+P_s^{(2)}(g_{a\gamma\gamma}^2,t).
\end{align}
In addition to the axion-induced signal, the intracavity power is also affected by factors such as pump-source amplitude noise, phase noise, and cavity thermal noise. We first discuss the effects of pump-source amplitude and phase noise on the intracavity signal power. The pump electric-field signal can be expressed as
\begin{align}
   \tilde{E}_0(t)=\tilde{E}_0(1+\epsilon(t))\cos{(\omega_0t+\phi(t))},
\end{align}
where $\epsilon(t)=\delta \tilde{E}_0/\tilde{E}_0$ represents the relative amplitude fluctuation of the pump electric field, and its power spectral density (PSD) is defined as $S_\epsilon(f)=\langle|\epsilon(f)|^2\rangle$. For pump amplitude noise, under the small-fluctuation approximation, the power fluctuation induced by the electric-field amplitude fluctuation can be written as
$\delta P_A=2P_s^{(0)}\epsilon(t)$.
Accordingly, its power spectral density is related to the relative intensity noise (RIN) as
${\rm RIN}(f)=S_{\delta P_A}(f)/(P_s^{(0)})^2=4S_\epsilon(f)$. The RIN characterizes the relative intensity noise level of the pump power as a function of frequency. In the absence of specific experimental noise parameters or device specifications for the pump source, we treat RIN as an external system parameter and adopt ${\rm RIN}=-120{\rm dB/Hz}$ as a reference value for the numerical calculations.
For a detection bandwidth of $\Delta f$, the variance of the power fluctuation induced by amplitude noise is given by
$\sigma_A^2=\int_{\Delta f}S_{\delta P_A}(f)df$.
When the RIN is approximately constant over the detection bandwidth, this expression can be further written as
$\sigma_A=P_s^{(0)}\sqrt{{\rm RIN}\Delta f}$,
where $\sigma_A$ denotes the RMS value of the power fluctuation induced by the pump amplitude noise.

On the other hand, $\phi(t)$ represents the phase fluctuation of the pump electric field, with its power spectral density defined as $S_\phi(f)=\langle|\phi(f)|^2\rangle$. We adopt a single-sideband phase-noise parameter of $\mathcal{L}=2\times10^{-10}{\rm rad^2/Hz}$ and approximate the phase-noise PSD as $S_\phi(f)=-100{\rm dB ~rad^2/Hz}$ over the relevant frequency range. Under the small-phase-fluctuation approximation, the induced intracavity power fluctuation is given by $\delta P_\phi(t)=\frac{\partial P_s^{(0)}}{\partial\phi}\phi(t)$, yielding $S_{\delta P_\phi}(f)=|\frac{\partial P_s^{(0)}}{\partial\phi}|^2S_\phi(f)$. Thus, within a detection bandwidth $\Delta f$, its RMS value is $\sigma_{P,\phi}=|\frac{\partial P_s^{(0)}}{\partial\phi}|\sqrt{S_\phi\Delta f}$.
 And shot noise $P_{\rm shot1}=\sqrt{P_s^{(0)}\hbar\omega_a/\tau}$ arising from photon-number fluctuations in the pump field. With their contributions integrated over the detection bandwidth $\Delta f$ to obtain the equivalent noise power entering the subsequent antenna coupling, transmission chain, and double-balanced homodyne detection system.

Next, we further consider the cavity thermal noise and the propagation of noise after the signal is coupled out through the antenna and transmitted through the transmission chain.

First, the resonant cavity itself exhibits thermal noise. In the classical limit, the cavity thermal-noise power can be expressed as
\begin{align}
P_{T,cav}&=k_BT_{cav}\Delta f.
\end{align}
Therefore, taking into account the pump-source amplitude noise, phase noise, shot noise, and cavity thermal noise, the total intracavity power can be expressed as
\begin{align}
P_{cav}=P_s(t)+\delta P_A+\delta P_\phi+P_{\rm shot1}+P_{T,cav}.
\end{align}
Subsequently, the cavity mode is coupled out through the antenna into the external measurement chain. At the antenna port, in addition to the intracavity signal coupled out through the antenna, the antenna and its external environment introduce additional equivalent thermal noise into the measurement system, denoted by $P_{N,\mathrm{ant}}$. Meanwhile, the external coupling quality factor $Q_i=\beta Q_c$, together with the quality factor of the cavity mode, determines the power extraction efficiency of the intracavity mode. Defining the antenna coupling efficiency as $\eta=\frac{\beta}{1+\beta}$, the total power at the antenna port can be expressed as
\begin{align}
    P_{ant}=\eta P_{cav}+P_{N,ant},
\end{align}
where $P_{N,\mathrm{ant}}$ denotes the equivalent noise power coupled into the measurement chain by the antenna and its external environment. In the classical limit, this noise power can be expressed as
\begin{align}
P_{T,ant}=k_BT_{ant,eq} \Delta f,
\end{align}
where $T_{\mathrm{ant,eq}}$ denotes the equivalent noise temperature at the antenna port.

After being coupled out through the antenna, the signal further enters the transmission chain. During transmission, the transmission line, connectors, and other components introduce power losses as well as additional thermal noise. Defining the power transmission coefficient of the transmission chain as $h<1$, the total power after transmission can be written as
\begin{align}
    P_{line}=hP_{ant}+P_{N,line},
\end{align}
where $P_{N,\mathrm{line}}$ denotes the equivalent noise power introduced by the transmission chain itself. Accordingly, this noise power can be expressed as
\begin{align}
P_{N,line}=k_BT_{\rm line ,eq}\Delta f,
\end{align}
where $T_{\mathrm{line,eq}}$ denotes the equivalent noise temperature of the transmission chain.

Therefore, before entering the double-balanced homodyne detection system, the total output power of the transmission chain can be further decomposed into the coherent signal power and the random noise power, namely,
\begin{align}
    P_{line}=P_{coh}+ P_N,
\end{align}
where the coherent power is primarily contributed by the intracavity signal and the amplitude and phase fluctuations of the pump source, and can be expressed as
\begin{align}
    P_{coh}=\eta h (P_s(t)+\delta P_A+\delta P_\phi+P_{\rm shot1}).
\end{align}
Defining $P_c=\eta hP_s^{(0)}[1+(P_s^{(1)}+P_s^{(2)})/P_s^{(0)}]$, the corresponding complex voltage signal can be expressed as
\begin{align}
    \tilde{V}_c=\sqrt{Z_0P_c}[1+\epsilon(t)]e^{i(\phi_c+\phi(t)},
\end{align}
where $Z_0$ denotes the characteristic impedance of the transmission line, and $\phi_c$ denotes the initial phase of the coherent signal (including the axion phase $\phi_a$). The corresponding coherent voltage signal is given by:
\begin{align}
    V_{c}(t)
=
\sqrt{2Z_0P_{c}}\,[1+\epsilon(t)]
\cos[\omega_0t+\phi_c+\phi(t)].
\end{align}

On the other hand, the system also contains a random noise voltage arising from the combined contributions of cavity thermal noise, antenna-port noise, and transmission-chain noise. This noise voltage can be expressed as
\begin{align}
    \tilde{V}_N=\tilde{V}_{T,cav}+\tilde{V}_{N,ant}+\tilde{V}_{N,line}=V_{N,I}+iV_{N,Q},
\end{align}
where $V_{N,I}$ and $V_{N,Q}$ denote the projections of the random noise voltage onto the two orthogonal $I$ and $Q$ components, respectively. Defining the random complex noise voltage such that $\langle|\tilde{V}_{N}|^2\rangle=Z_0P_N$, the two statistically independent orthogonal noise components with equal variances satisfy
$\langle|\tilde{V}_{N,I}|^2\rangle=\langle|\tilde{V}_{N,Q}|^2\rangle=\langle|\tilde{V}_{N}|^2\rangle/2$
The random noise power is given by:
\begin{align}
     P_{N}
=
\eta hP_{T,\rm cav}
+hP_{N,\rm ant}
+P_{N,\rm line}.
\end{align}
Assuming that all noise sources are statistically independent, we have
\begin{align}
    \tilde{V}_{N}=\tilde{V}_{T,\rm cav}
+
\tilde{V}_{N,\rm ant}
+
\tilde{V}_{N,\rm line},
\end{align}
where the mean-square voltages of the respective noise components satisfy
$\left\langle|\tilde{V}_{T,\rm cav}|^2\right\rangle
=Z_0\eta hP_{T,\rm cav},\
\left\langle|\tilde{V}_{N,\rm ant}|^2\right\rangle
=Z_0hP_{N,\rm ant},\
\left\langle|\tilde{V}_{N,\rm line}|^2\right\rangle
=Z_0P_{N,\rm line}.
$

Therefore, the total complex voltage entering the double-balanced homodyne detection system can be expressed as $\tilde{V}=\tilde{V}_c+\tilde{V}_N$. Transforming it into the time domain, the input signal voltage can be written as
\begin{align}
    V=&\sqrt{2Z_0P_{c}}\,[1+\epsilon(t)]
\cos[\omega_0t+\phi_c+\phi(t)]\nonumber\\
&+\sqrt{2}[V_{N,I}(t)\cos{(\omega_0t)}-V_{N,1}(t)\sin{(\omega_0t)}]
\end{align}
where the first term represents the coherent signal containing the pump amplitude and phase fluctuations, while the second term represents the two orthogonal $I$ and $Q$ components of the random noise voltage around the carrier frequency $\omega_0$.

For the double-balanced homodyne detection system, the local oscillator (LO) signal entering the quadrature mixer can be represented by the complex voltage $\tilde{V}_L=\sqrt{Z_0P_L}e^{i\phi_L}$, where $P_L$ denotes the LO power and $\phi_L$ denotes the phase of the LO signal. The corresponding time-domain voltage of the $I$-channel LO is given by
\begin{align}
    V_{LI}(t)=\sqrt{2Z_0P_L}\cos{(\omega_0t+\phi_L)}
\end{align}
Thus, the input signal and the LO signal are fed into the double-balanced homodyne detection system. For the $I$-channel, the input signal voltage is combined with the LO signal through the sum and difference operations, yielding
\begin{align}
    V_{I+}(t)&=\sqrt{2Z_0P_{c}}\,[1+\epsilon(t)]
\cos[\omega_0t+\phi_c+\phi(t)]\nonumber\\
&+\sqrt{2}V_{N,I}(t)\cos{(\omega_0t)}+\sqrt{2Z_0P_L}\cos{(\omega_0t+\phi_L)}\nonumber\\
 V_{I-}(t)&=\sqrt{2Z_0P_{c}}\,[1+\epsilon(t)]
\cos[\omega_0t+\phi_c+\phi(t)]\nonumber\\
&+\sqrt{2}V_{N,I}(t)\cos{(\omega_0t)}-\sqrt{2Z_0P_L}\cos{(\omega_0t+\phi_L)}    
\end{align}
The two signals are then fed into separate square-law detectors. Let the detector responsibility be $\mathcal{R}$, with units of $\rm V/W$. After square-law detection and low-pass filtering, the two output voltages of the $I$-channel are given by
\begin{align}
    V_{LPI+}=&\mathcal{R}\langle\frac{V_{I+}^2}{Z_0}\rangle=\frac{1}{2Z_0}\mathcal{R}[V_c^2+V_{N,I}^2+V_{LI}^2\nonumber\\
    &+2(V_cV_{N,I}+V_{N,I}V_{LI}+V_cV_{LI})]\nonumber\\
    V_{LPI-}=&\mathcal{R}\langle\frac{V_{I-}^2}{Z_0}\rangle=\frac{1}{2Z_0}\mathcal{R}[V_c^2+V_{N,I}^2+V_{LI}^2\nonumber\\
    &+2(V_cV_{N,I}-V_{N,I}V_{LI}-V_cV_{LI})].
\end{align}
Taking the difference between the two output signals of the $I$-channel eliminates the common power term that is independent of the LO phase, yielding
\begin{align}
    V_I=V_{LPI+}-V_{LPI-}=\frac{2\mathcal{R}}{Z_0}(V_{N,I}V_{LI}+V_cV_{LI})=V_{I,c}+V_{I,N},
\end{align}
where $V_{I,N}={2\mathcal{R}}V_{N,I}V_{LI}/{Z_0}$ represents the base-band output component associated with the random noise in the $I$-channel, while
\begin{align}
    V_{I,c}=2\mathcal{R}\sqrt{2P_cP_L}[1+\epsilon(t)]\cos{(\Delta \phi(t))},
\end{align}
represents the coherent signal output of the $I$-channel, where $\Delta\phi(t)=\phi_c+\phi(t)-\phi_L=\Delta\phi_0+\phi(t)$. Since $\phi(t)\sim 0$, under the small-phase-fluctuation approximation, $\phi(t)\ll1$, we have
\begin{align}
    V_{I,c}=2\mathcal{R}\sqrt{2P_cP_L}[\cos{(\Delta \phi_0)}+\epsilon(t)\cos{(\Delta \phi_0)}-\phi(t)\sin{(\Delta \phi_0)}].
\end{align}

Similarly, for the $Q$-channel, the time-domain voltage of the LO signal is given by
\begin{align}
    V_{LQ}=\sqrt{2Z_0P_L}\sin{(\omega_0t+\phi_L)},
\end{align}
The input signal $V(t)$ is then fed into the sum and difference ports of the $Q$-channel, where it is combined with the LO signal $V_{LQ}$, yielding
\begin{align}
    V_{Q+}(t)&=\sqrt{2Z_0P_{c}}\,[1+\epsilon(t)]
\cos[\omega_0t+\phi_c+\phi(t)]\nonumber\\
&-\sqrt{2}V_{N,Q}(t)\cos{(\omega_0t)}+\sqrt{2Z_0P_L}\sin{(\omega_0t+\phi_L)}\nonumber\\
 V_{Q-}(t)&=\sqrt{2Z_0P_{c}}\,[1+\epsilon(t)]
\cos[\omega_0t+\phi_c+\phi(t)]\nonumber\\
&-\sqrt{2}V_{N,Q}(t)\cos{(\omega_0t)}-\sqrt{2Z_0P_L}\sin{(\omega_0t+\phi_L)}  
\end{align}
The two signals are then passed through separate square-law detectors and low-pass filters, yielding
\begin{align}
    V_{LPQ+}=&\mathcal{R}\langle\frac{V_{Q+}^2}{Z_0}\rangle=\frac{1}{2Z_0}\mathcal{R}[V_c^2+V_{N,Q}^2+V_{LQ}^2\nonumber\\
    &+2(-V_cV_{N,Q}-V_{N,Q}V_{LQ}+V_cV_{LQ})]\nonumber\\
    V_{LPQ-}=&\mathcal{R}\langle\frac{V_{Q-}^2}{Z_0}\rangle=\frac{1}{2Z_0}\mathcal{R}[V_c^2+V_{N,Q}^2+V_{LQ}^2\nonumber\\
    &+2(-V_cV_{N,Q}+V_{N,Q}V_{LQ}-V_cV_{LQ})]
\end{align}
Taking the difference between the two output signals of the $Q$-channel, we obtain
\begin{align}
      V_Q=V_{LPQ+}-V_{LPQ-}=\frac{2\mathcal{R}}{Z_0}(-V_{N,Q}V_{LQ}+V_cV_{LQ})=V_{Q,c}+V_{Q,N},
\end{align}
where $V_{Q,N}=-{2\mathcal{R}}V_{N,Q}V_{LQ}/{Z_0}$ represents the baseband output component associated with the random noise in the $Q$-channel. Under the small-phase-signal approximation, the coherent signal output of the $Q$-channel is given by:
\begin{align}
    V_{Q,c}=2\mathcal{R}\sqrt{2P_cP_L}[\sin{(\Delta \phi_0)}+\epsilon(t)\sin{(\Delta \phi_0)}+\phi(t)\cos{(\Delta \phi_0)}].
\end{align}
For simplicity, we define the system gain factor as $G=2\mathcal{R}\sqrt{2P_cP_L}$. Therefore, the total output power corresponding to the $I$- and $Q$-channels of the double-balanced homodyne detection system can be expressed as
\begin{align}
    P_{out}=&\frac{(V_{I,c}^2+V_{Q,c}^2)}{Z_D}=\frac{1}{Z_D}[G^2(1+\epsilon(t))^2+G^2\phi^2(t)+V_{I,N}^2+V_{Q,N}^2]\nonumber\\
    &+2G(1+\epsilon(t))\frac{1}{Z_D}[V_{IN}\cos(\Delta \phi_0)+V_{QN}\sin(\Delta \phi_0)]\nonumber\\
    &-2G\phi(t)\frac{1}{Z_D}[V_{IN}\sin(\Delta \phi_0)-V_{QN}\cos(\Delta \phi_0)],
\end{align}
where the random noise voltage satisfies $V_{N,I}^2+V_{N,Q}^2=\langle|\tilde{V}*N|^2\rangle=Z_0P_N$. Therefore, the noise output power generated by the random noise in the $I$- and $Q$-channels after square-law detection satisfies $V^2_{I,N}+V_{Q,N}^2=8\mathcal{R}^2P_NP_L$. If we further assume that the random noise $P_N$ is statistically independent of the pump-source amplitude noise $\epsilon(t)$ and phase noise $\phi(t)$, the corresponding cross terms vanish upon statistical averaging. Thus, the average output power can be simplified as
\begin{align}
    P_{out}=&\frac{(V_{I,c}^2+V_{Q,c}^2)}{Z_D}=\frac{1}{Z_D}[G^2[(1+\epsilon(t))^2+\phi^2(t)]+V_{IN}^2+V_{QN}^2].
\end{align}
In the absence of an axion signal, i.e., considering only the background pump power $P_s^{(0)}$, the system output power can be expressed as
\begin{align}
    P_{out}^{n}=\frac{8\eta h\mathcal{R}^2P_LP_s^{(0)}}{Z_D}[(1+\epsilon(t))^2+\phi^2(t)]+\frac{1}{Z_D}8\mathcal{R}^2P_NP_L.
\end{align}
In the presence of an axion-induced signal, the system output power is given by
\begin{align}
    P_{out}^{n}=&\frac{8\eta h\mathcal{R}^2P_L}{Z_D}(P_s^{(0)}+P_s^{(1)}+P_s^{(2)})[(1+\epsilon(t))^2+\phi^2(t)]\nonumber\\
    &+\frac{1}{Z_D}8\mathcal{R}^2P_NP_L.
\end{align}
Thus, the final output power of the double-balanced homodyne detection system can be expressed as a detection observable comprising contributions from the coherent signal, pump-source amplitude and phase noise, and random noise. This provides the basis for constructing the detection estimator for a given detection bandwidth $\Delta f$ and integration time $t_{\rm in}$, and subsequently analyzing the statistical SNR of the system using the Dicke radiometer equation.

Within the framework of the Dicke radiometer equation, the system signal-to-noise ratio (SNR) can be expressed as:
\begin{align}
    {\rm SNR_{Dicke}}=\frac{P_{out}^{s}}{\sigma_{N,avg}}=\frac{P_{out}^s\sqrt{t_{in}\Delta f }}{\sigma_N},
\end{align}
where $\sigma_{N,\mathrm{avg}}=\sigma_N/\sqrt{t_{\rm in}\Delta f}$, where $t_{\rm in}$ denotes the total integration time and $\Delta f$ denotes the effective detection bandwidth for a single measurement. Here, $P_{\rm out}^{s}$ represents the equivalent signal power of the axion-induced signal after passing through the cavity, antenna, transmission chain, and double-balanced homodyne detection system, namely, $P_{\rm out}^{s}=\frac{8\eta h\mathcal{R}^2P_L}{Z_D}(P_s^{(1)}+P_s^{(2)})$. Assuming that all noise sources are statistically independent, the total system noise can be expressed as the combined contributions from random noise, pump-source amplitude noise, pump-source phase noise and shot noise i.e.,
\begin{align}
    \sigma_N=\sqrt{\sigma_{P_N}^2+\sigma_{A1}^2+\sigma_{P,\phi 1}^2+\sigma_{\rm shot1}^2},
\end{align}
where $\sigma_{A1}$
  and $\sigma_{P,\phi 1}$
  represent the root-mean-square (RMS) output noise values corresponding to the pump-source amplitude noise and phase noise, respectively, after double-balanced homodyne detection. Meanwhile, $\sigma_N$  and $\sigma_{\rm shot1}$ denote the RMS output noise values of the equivalent thermal noise from the cavity and transmission chain and the shot noise after double-balanced homodyne detection, respectively.  Based on the signal and noise propagation described above, the system noise can be approximately expressed as
\begin{align}
\sigma_N
\simeq&
\frac{8\mathcal{R}^2P_L}{Z_D}
\left(
\eta hP_{T,\rm cav}
+hP_{N,\rm ant}
+P_{N,\rm line}
\right)
\nonumber\\
&+
\frac{8\eta h\mathcal{R}^2P_L}{Z_D}
\left(
\sigma_A+\sigma_{P,\phi}+P_{\rm shot1}
\right).
\end{align}
Substituting the above results into the Dicke SNR expression, we obtain
\begin{align}
    {\rm SNR_{Dicke}}=\frac{\eta h(P_s
    ^{(1)}+P_s
    ^{(2)})\sqrt{t_{in}\Delta f}}{\eta hP_{T,cav}+hP_{N,ant}+P_{N,line}+\eta h(\sigma_A+\sigma_{P,\phi}+P_{\rm shot1})}
\end{align}
For the antenna-coupling noise and transmission-line noise, these contributions can be further treated as equivalent thermal noise and described in terms of an equivalent noise temperature. Assuming that the system is referenced to the cavity temperature $T_{\rm cav}$, and that each additional noise-contributing stage increases the equivalent noise temperature by $x,{\rm K}$, the above SNR can be further written as
\begin{align}
    {\rm SNR_{Dicke}}=\frac{(P_s
    ^{(1)}+P_s
    ^{(2)})\sqrt{t_{in}\Delta f}}{k_B\Delta f[(1+\frac{1}{\eta}+\frac{1}{\eta h})T_{cav}+(\frac{x}{\eta}+\frac{2x}{\eta h})]+\sigma_A+\sigma_{P,\phi}+P_{\rm shot1}}.
\end{align}
According to the pump-source noise model described above, the RMS power fluctuations due to amplitude and phase noise are $\sigma_A=P_s^{(0)}\sqrt{{\rm RIN}\Delta f}$ and $\sigma_{P,\phi}=P_s^{(0)}\sqrt{S_\phi\Delta f}$, respectively. For the parameter range considered in this work, taking $\tilde{{B}}_{\rm 0,sys}=1{\rm nT},P_s^{(0)}\sim10^{-8}\rm W$, $\Delta f=100\rm Hz$, ${\rm RIN}=-120\rm dB/Hz$, and $S_\phi\sim10^{-10}\rm rad^2/Hz$, we obtain $\sigma_A\sim10^{-13}\rm W$ and $\sigma_{P,\phi}\sim10^{-12}\rm W$, respectively. In comparison, even at room temperature, the cavity thermal-noise power is only about $10^{-18}\rm W$, which is several orders of magnitude smaller than the pump-source amplitude and phase noise. Therefore, under the parameters adopted in this work, the system noise is dominated by the pump-source amplitude and phase noise, while the contributions from cavity thermal noise, antenna noise, and transmission-line thermal noise to the final SNR can be neglected to a good approximation.

We further consider the power of the axion-induced signal. According to the mode-coupling relations discussed above, the zeroth- and first-order signal powers can be written as $P_s^{(0)}=\gamma_m\int_V\varepsilon_0 |\tilde{E}_0|^2dV$ and $P_s^{(1)}=\gamma_m\int_V\varepsilon_0 |\tilde{E}_0||E^{(1)}|dV$, respectively, while the second-order signal power is $P_s^{(2)}=\gamma_m\int_V\varepsilon_0 |E^{(1)}|^2dV$. Therefore, under the same normalization condition, we have $P_s^{(1)}=\sqrt{P_s^{(0)}P_s^{(2)}}$. Since $P_s^{(0)}\gg P_s^{(2)}$ under the pump conditions considered in this work, the second-order term $P_s^{(2)}$ can be neglected in the signal contribution, yielding
\begin{align}
     {\rm SNR_{Dicke}}&=\frac{P_s
    ^{(1)}\sqrt{t_{in}\Delta f}}{P_s^{(0)}\sqrt{{\rm RIN}\Delta f}+P_s^{(0)}\sqrt{S_\phi \Delta f}+P_{\rm shot1}}\nonumber\\
    &=\frac{\sqrt{P_s^{(2)}t_{in}\Delta f}}{\sqrt{P_s^{(0)}}(\sqrt{{\rm RIN}\Delta f}+\sqrt{S_\phi \Delta f})+\sqrt{\hbar\omega_a/\tau}},
\end{align}
When the system reaches the detection threshold, ${\rm SNR}_{\rm Dicke}\geq1$, the minimum detectable axion–photon coupling can be obtained as
\begin{align}
    g_{a\gamma\gamma}^{(1)}\ge\frac{\sqrt{P_s^{(0)}}(\sqrt{{\rm RIN}\Delta f}+\sqrt{S_\phi \Delta f})+\sqrt{\hbar\omega_a/\tau}}{\sqrt{\kappa t_{in}\Delta f}},
\end{align}
where $\kappa=P_{s}^{(2)}/g_{a\gamma\gamma}^2, ~P_{s}^{(2)}=g_{a\gamma\gamma}^2a_0^2\omega_aQ\bar{B}^2 C_1V/8\mu_0$.

It is evident that, under the present noise model and parameter conditions, the detection sensitivity of the system is primarily limited by the pump-source amplitude, phase noise and shot noise, while the cavity thermal noise, antenna noise, and transmission-line thermal noise are all substantially smaller than the pump-source noise. 



\end{document}